# How to find MH370?

Martin Kristensen

(*Department of Engineering, Aarhus University, Finlandsgade 22, DK-8200 Aarhus N, Denmark*)
(Email: mk@eng.au.dk)

The disappearance of flight MH370 is possibly the greatest mystery in aviation history. A large zone in the Southern Indian Ocean was searched unsuccessfully leaving an open case and an unacceptable situation for the family members. We discuss the scientific difficulties with locating the plane through satellite data and develop an improved analysis using least square curve fitting of analytical non-Euclidean route equations providing robust topology-optimization with perturbation theory handling satellite movement. We find four independent solutions with the final part of the flight following a great circle. Two are located in stable minima for the error-function, and two unstable ones agree poorly with most data. One stable solution coincides with the Inmarsat-result, but fails to explain additional data. Our best solution leads to an entirely different location agreeing with other data from debris, acoustics, an eyewitness report, the received microwave power, several contrails, two seismic detectors at Christmas Island, and with a mushroom impact cloud providing the final proof where to find the rest of the airplane and the black boxes.

1. INTRODUCTION. After contact was lost with flight MH370 on the route from Kuala Lumpur to Beijing on March 8, 2014 the only available data concerning its route came from a military radar facility in Malaysia (News, 2017) and a number of handshakes with a satellite from Inmarsat (Ashton et al., 2014). Radar data tells us that after shutting down transponders and communication systems it made a sharp turn and flew along the border to Thailand, past Penang to the waypoint MEKAR in the Malacca Strait where military radar lost contact (News, 2017). Then the satellite handshakes restarted after interruption of the satellite unit (SDU), possibly due to a power failure. The handshakes provide accurate time differences and thereby distances to the satellite 3F1 at approximately geostationary location west of the Maldives, represented by so-called Burst Time Offset (BTO) (Ashton et al., 2014). The sum of the Doppler shifts for the communication loop was also measured and the local Doppler shift from the airplane movement deduced through the Burst Frequency Offset (BFO). Since these represent frequency shifts due to the radial projection of the movement of the plane, measured simultaneously with BTO at roughly one-hour intervals after the reboot, and two data points due to attempted phone calls, it is possible to deduce some route information. However, the accuracy of the Doppler shifts (1% of the maximum value) was believed insufficient to determine a precise route.

If the satellite were perfectly geostationary we would have been practically stuck here, since we could only conclude that it reached a certain distance from the last BTO, giving roughly a circle called the seventh arc after numbering the handshakes. This corresponds to 2662 nmi (nautical mile (nmi) = 1.852 km) along a spherical earth surface from the ground projection of 3F1. This arc is cut around 55% shorter due to the maximum fuel range of the plane but still represents an unsurmountable difficulty to search, particularly when allowing for moderate movement from the last handshake to the end. In other words, the plane effectively disappeared. Physically this is easy to understand, since the satellite measurements only provide radial information, but no direction.

Fortunately the satellite is not completely ideal and wanders, predominantly in north-south direction (Ashton et al., 2014), and the maximum speed and minimum turning radius for the plane also give some guidance. During most of the flight the satellite moved south after reaching its northern extremum roughly an hour after the SDU restarted. This opens the possibility to distinguish between northern and southern flight routes since the perturbation to the relative airplane velocity (for SDU-satellite Doppler shift) has opposite sign for northern and southern routes. For routes going straight north or south, this gives a 10% difference. This is ten times the measurement uncertainty for the BFO leaving no doubt routes going straight north are inconsistent with the data, as shown by Ashton (2014). Routes towards northeast are even worse. However, in north-western direction there is a chance of a route with poor fit and range near the acceptance limit because the east-west satellite movement mixes with the north-south movement relaxing fitting conditions. We initially ignore this type of solution since it is unlikely that any airplane could penetrate radar surveillance by several countries including India and China on a north-western route. In addition, it would have to fly at relatively low speed along a curved route to fit the data, and thereby be forced to, but at the same time practically unable to, pass the tall mountains in Himalaya. In addition, some of the phones belonging to the passengers would likely have been on and left electronic footprints by handshakes with the Chinese network. Finally, all debris would have to be planted. Later we develop a simple test to find all remaining solutions and settle the issue rigorously. This leads us to conclude that only one of the solutions is correct, and define a new and dramatically smaller search zone near Christmas Island. This also explains why all searching until now was unsuccessful.

2. ANALYSIS OF SATELLITE DATA. Soon after military radar contact was lost, the SDU restarted unexpectedly. After some time without power, it was cold, so it rebooted and reheated simultaneously leading to some highly unreliable handshakes (BTO/BFO) due to temperature drift. These points must be discarded. Around 18:28 UTC, the measurement quality was improving in agreement with results from previous SDU cold-starts (Holland, 2017). At 18:40, Malaysia Airlines called the plane via satellite phone. Nobody answered, but the call provided BFO values and defined the starting point for the hourly handshakes. Unfortunately, telephone calls give no BTO (Ashton et al., 2014). Since the airplane had moved from MEKAR so its height, speed and course were no longer known independently from radar data, it is impossible to extract precise knowledge from these two points without making assumptions or having additional knowledge. It is therefore not until the first regular handshake at 19:41 that complete and precise data is available for systematic analysis.

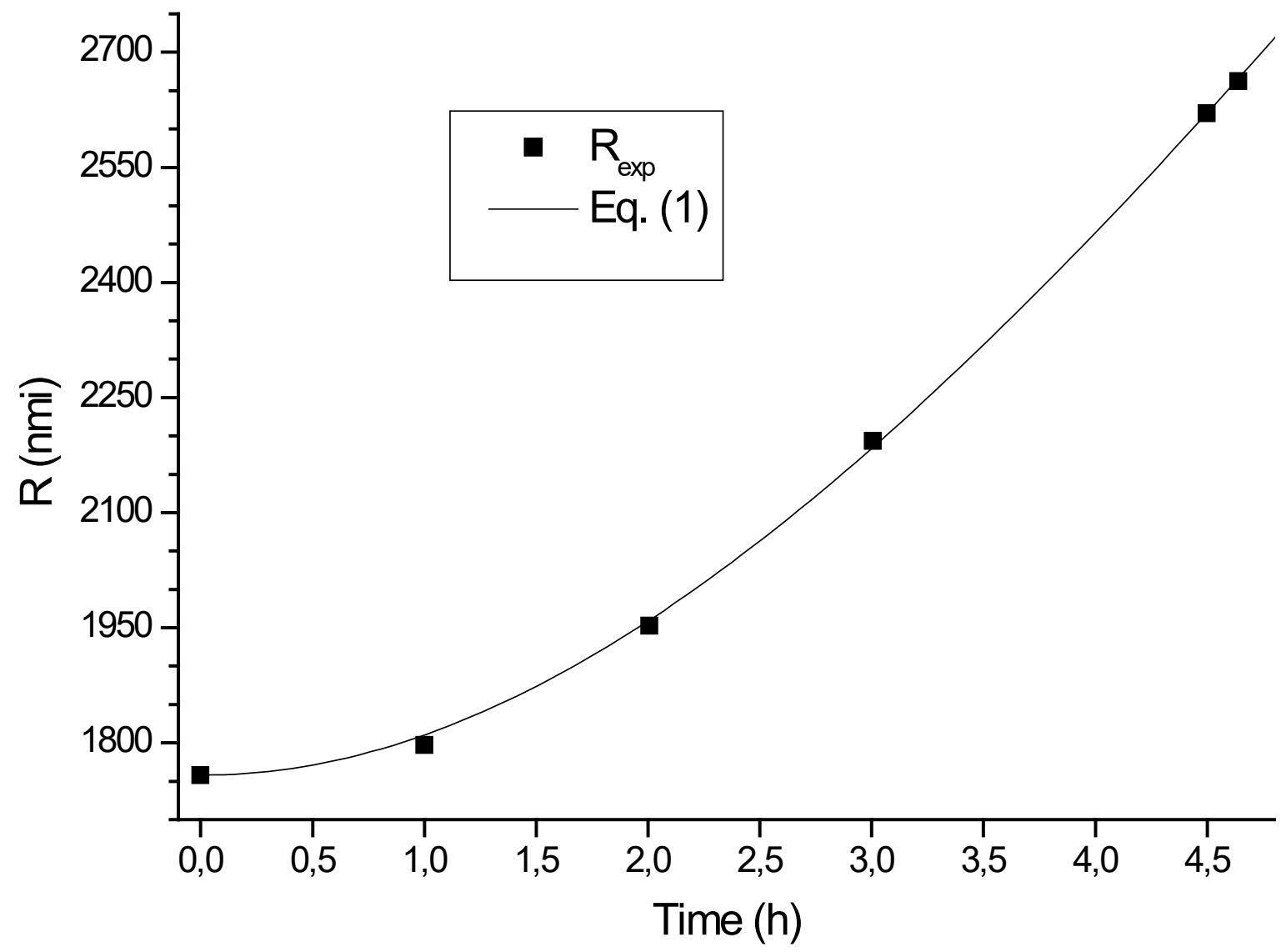

Figure 1. Equation (1) compared to the radial distances $R = R_{exp}$ on a spherical earth.

At 18:28, there are two qualitative options: A northern or a southern route. The handshake at 18:28 allows for these two solutions to a quadratic equation. Concerning the route after 18:28 the BFO at 18:40 provides some selection among the possibilities but unfortunately the problem is underdetermined and the data too inaccurate to distinguish between north and south. Importantly the rest of the points look like they are located on a straight line with constant speed when considering the BTO values. If one converts the BTO values to distances, $R$, along the earth surface from the satellite ground projection using equations (3) and (4), these are modelled well by a simple equation describing a straight line on a flat earth with an exclusively transverse speed of $v$ = 800 km/h, starting at $t_1$ (19:41)

$$R = 1758nmi\sqrt{1 + \left(0.2457(t - t_1)\right)^2} \tag{1}$$

where the front factor is $R$ at $t_1$ = 19:41 and the value 0.2457 is the ratio of $v$ and $R(t_1)$. This is a strong lead, in particular since the agreement with measured values is good as shown in figure 1. Furthermore, the BFO values also agree with expectations from such a straight flight, but with less precision, and with the exception of 19:41, which has movement towards the satellite rather than away from it as the rest of the points, compared to zero movement as equation (1) predicts. In fact, it is impossible to reconcile the BFO and BTO information at 19:41 without assuming one of the following:

1) The point at 19:41 is not at all on the same straight line as the rest.
2) There is roughly 10-degree heading change between 19:41 and 20:41.
3) The minimum distance to the satellite lies between 19:41 and 20:41.
4) There is a U-turn between 19:41 and 20:41.

We combined BTO and BFO information in a detailed mathematical analysis to gain deeper insight and found the sign-change for the aircraft Doppler-shift of key importance to select the correct option.

We eliminate two of the four possibilities by initially considering a flat earth, and improve by a Taylor expansion of the spherical solution. For a flat earth the surface angle towards the satellite ground projection, $\theta$, is linked to the speed, $v$, and radius along the earth surface from the ground projection, $R$, by

$$1.1 \cdot \frac{R}{{R_J}^2} \cdot v \cos\theta = 1.1 \cdot \left(\frac{v}{R_J}\right)^2 \cdot (t - t_0) \qquad (2)$$

where $R_J$ is the earth radius and $v$ the airplane speed. Using that $v \cdot \cos(\theta)$ is proportional to the Doppler shift we plot this relationship in figure 2 against time, $t$, measured from 19:41 using the BFO values converted to Doppler shifts (Ashton et al., 2014) and the $R$-values. We find a relatively good linear fit for all the later points but a poor one for 19:41, and moderate deviation for 20:41. By replacing the simplified flat earth version (2) (only valid for short distances) with a spherical solution derived from equations (7) and (8) and Taylor expanded to second order in the point at 22:41 (the best approximation which is still a straight line, and labelled equation 2A) one finds a nice fit from 20:41 as shown in the same figure. We multiplied equation (2) by 1.1 to match the simple equation with the optimally Taylor-expanded Doppler shifts at short distances. A linear fit gives $v = 784$ km/h using the Taylor formula. As illustrated, the point at 19:41 deviates by a Doppler sign change (indicating U-turn) and is therefore not included in the fit. Instead, we find the natural zero-point, $t_0$, at -0.61 hours (before 19:41) eliminating option 3, since the minimum satellite distance, $R_{min}$, and corresponding sign change must occur at $t_0$. Agreement with equation (1) and sign change at $t = 0$ makes option 1 impossible. This leaves us only with options 2 or 4 (most likely 4). These lead to solutions in the Southern Indian Ocean and the Eastern Indian Ocean respectively, while the Iannello solution (Iannello and Godfrey, 2016) can connect to both options.

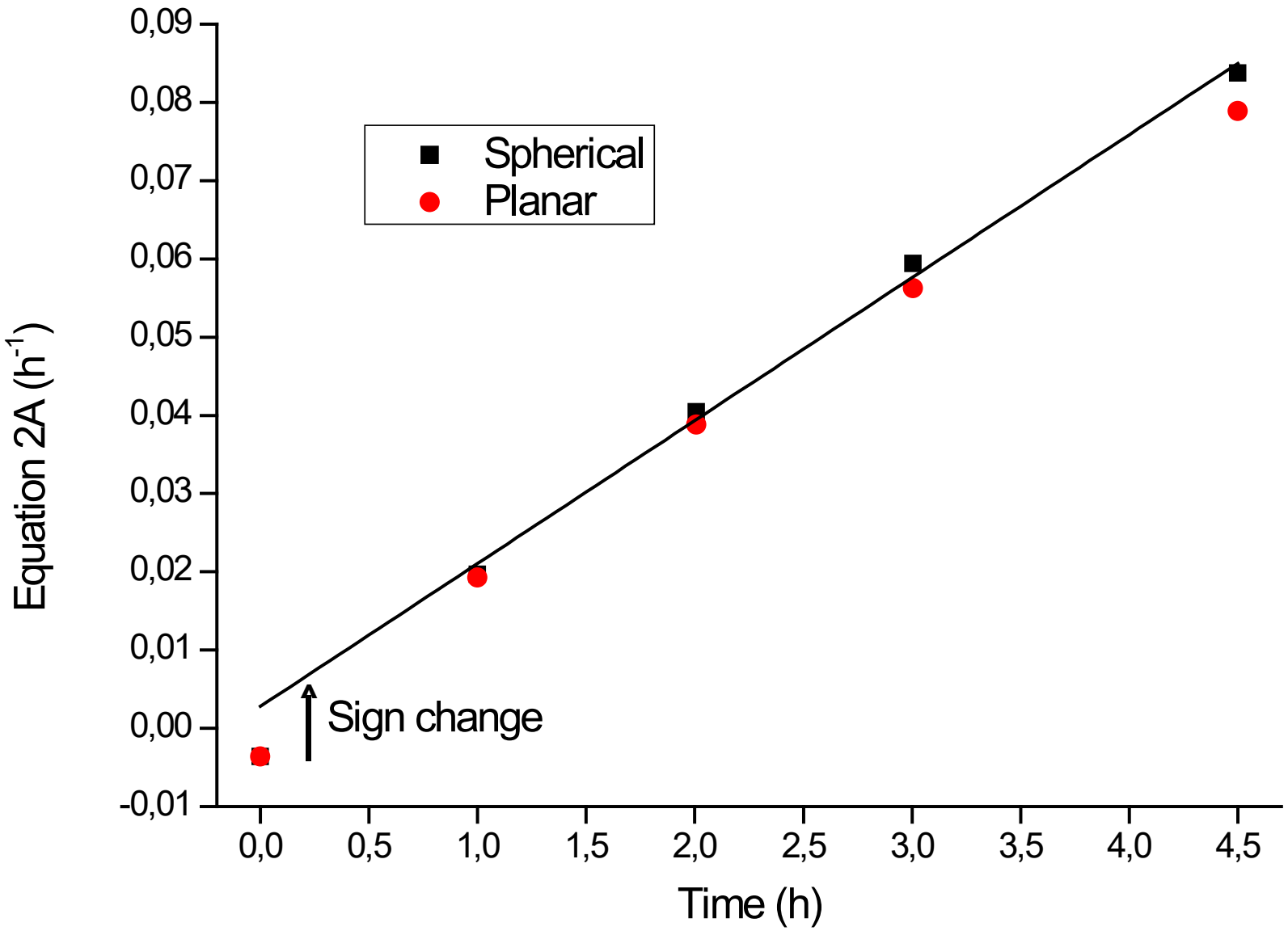

Figure 2. Data points from equation (2) using a flat earth description (red circles), and a second-order Taylor expansion (2A) of the spherical equation at 22:41 (black squares) providing the most accurate linear description for all distances, plotted against time after 19:41 and fitted with a straight line. It illustrates that the point at 19:41 would fit much better if the Doppler-effect changes sign due to U-turn soon after 19:41.

After establishing this overview of possible solutions, we perform the detailed analysis of the rest of the data. The analytical connection between the satellite position, the BTO value and the radius, $R$, along the earth surface from the satellite ground projection to the airplane is given by Ashton (2014) and Steel (2016)

$$R_{sat} = \sqrt{\left(R_J + H\right)^2 + \left(R_J + h\right)^2 - 2\left(R_J + H\right)\left(R_J + h\right)\cos\left(\frac{R}{R_J}\right)} \tag{3}$$

$$R_{sat} = \frac{c}{2}(BTO - bias) - R_{sat-Perth} \tag{4}$$

where $R_{sat}$ is the line-of-sight distance from the satellite to the airplane, $R_J = 6378$ km is the earth radius, $H$ is the satellite height above earth, $h$ is the flying height, $c$ is the speed of light, *bias* is an internal Inmarsat parameter and $R_{sat-Perth}$ the distance from the satellite to the relay station. We calculate $R$ for a spherical earth model to simplify the non-Euclidean equations. Finally, we make oblate projections (Wikipedia-1) to place them on the true earth. Please note that the $R$-values do not agree with a table originally published in the blog of (Steel, 2016) and later deleted. According to our calculations those values had up to one percent error.

All Doppler-shifts are classical, since it is safe to ignore relativistic contributions, atmospheric influence and gravitational shifts, which are all at least 30 times below the measurement uncertainty. We calculate the Doppler-shift from the change in the line-of-sight distance between the airplane and the satellite by differentiation of equation (3), initially ignoring the satellite earth-projection movement

$$\frac{\partial R_{sat}}{\partial t} = \frac{\partial H}{\partial t} + \left(1 + \frac{R_J}{R_J + H}\cos\left(\frac{R}{R_J}\right)\right) \cdot \sin\left(\frac{R}{R_J}\right) \cdot v\cos\theta \tag{5}$$

where $\theta$ is the angle between $R$ and the flight direction. We use $h = 11$ km, ignore height changes until the seventh arc, and assume $H >> R_J >> h$. This flight height was chosen after optimizing the entire problem (Plougmann and Kristensen, 2004) with $h$ fixed at 10, 11 and 12 km respectively, and finding slightly better fit for both stable solutions at 11 km (most pronounced south-easterly). In addition to the expected Doppler-shift due to the airplane velocity with respect to the satellite there is a height contribution from the satellite. It is unclear how Inmarsat handled this contribution. They split up the Doppler shift into two contributions: One entirely due to the satellite and one entirely due to the plane. Strictly speaking, this is wrong since Doppler-shift is a relative effect as pointed out by (Einstein, 1905), since the Doppler effect depends only on the relative motion of source and receiver. One can circumvent this by choosing a suitable inertial-system as reference. However, Inmarsat has not made a well-defined choice of inertial-system. If we calculate the contribution from the satellite ground-projection movement relative to its most northern point, (not an inertial system because of earth rotation) we get agreement with their satellite contribution except for last decimal round-off and the point at 21:41 where there is a somewhat larger deviation. This issue cannot be ignored since our model is so accurate that even the last decimal matters, and the deviation at 21:41 is more than $2\sigma$ for the BFO for other routes than Inmarsat's route where everything fits. We tried to add the satellite height change, but this makes agreement worse. Therefore, we have chosen to ignore the height-change and add its maximum size linearly to the error-bar leading to a doubling of $\sigma$ for the BFO. This only moves the predicted crash site a few km, so we can live with it. Under these conditions and using equation (5) the airplane Doppler shift, $\Delta\nu$, is

$$\Delta\nu = -\frac{\nu_0}{c}\left(1 + \frac{R_J}{R_J + H}\cos\left(\frac{\widetilde{R_n}}{R_J}\right)\right)\sin\left(\frac{\widetilde{R_n}}{R_J}\right) \cdot v\cos\left(\widetilde{\theta_n}\right) \tag{6}$$

where $\nu_0 = 1646652500$ Hz is the communication frequency (Ashton et al., 2014), and experimental numbers including perturbation effects from satellite movement are symbolized with a tilde over the variables ($\widetilde{R}$, $\tilde{\theta}$). We think using a detailed recipe from Inmarsat the $\sigma$ enlargement can be avoided and the agreement will be better for 21:41. The published Inmarsat procedure is perfect for their specific route but potentially problematic for other routes. However, an alternative possibility is that the plane passed a thunderstorm at 21:41.

Table 1. First order perturbations from the satellite movement on $R$, $v$ (along $R$) and $\theta$ for the two solutions towards 13.3˚S and 34.6˚S using values from (Ashton et al., 2014) and our own recalculations.

| Satellite perturbation | Solution towards 13.3˚S | | | Solution towards 34.6˚S | | |
|---|---|---|---|---|---|---|
| Time (UTC) | $-\Delta R$ (nmi) | $-v_R$ (km/h) | $\Delta\theta$ (°) | $-\Delta R$ (nmi) | $-v_R$ (km/h) | $\Delta\theta$ (°) |
| 20:41:05 | -1.54 | -3.94 | 0.113 | 0.6 | 3 | 0.02 |
| 21:41:27 | -1.72 | -2.09 | 0.410 | 6 | 12 | 0.19 |
| 22:41:22 | 2.48 | 3.85 | 0.785 | 18 | 21 | 0.47 |
| 00:11:00 | 18.71 | 14.48 | 1.304 | 48 | 34 | 1.02 |

After calculating the values of $R$ and $\Delta v$ we take the perturbations from the satellite ground-projection movement into account. We use classical first order perturbation theory (Stewart, 1990) to find the projection on the individual $R$-values and satellite Doppler effect for two solutions going south and southeast in table 1. By subtracting the perturbations from the measured values, we get the unperturbed result for a spherical earth in case the direction is correct. After a couple of tries, we found that we only arrive at self-consistent routes with good fits in two directions. In addition we found indication of a third solution between them, but its error ($\chi^2$) minimum is too shallow for a stable fit as shown in figure 3. Starting at any other direction leads to an iterative convergence towards one of the two stable solutions. South of the shallow minimum converging to the southern solution and north of it to the south-eastern solution, both after 3-4 iterations where the satellite perturbations are recalculated for each iteration.

Table 2. Fitting results with equations (6-8) for the solutions towards 13.3˚S and 34.6˚S with $\chi^2_{V.I.}$ = 31.7 nmi² and $\sigma_{Doppler}$ = 15 Hz, giving $\chi^2_{statistical}$ = 1.25 nmi².

| Route | $t_0$ (h) | $v$ (km/h) | $R_{min}$ (nmi) | $\chi^2_R$ (nmi²) | $\chi^2$ (nmi²) | $\widetilde{\theta_6}$ (°) |
|---|---|---|---|---|---|---|
| 13.3˚S | -0.4709±0.0035 | 796.87±0.50 | 1661.6±1.2 | 0.475 | 0.845 | 43.294±0.030 |
| 34.6˚S | -0.3800±0.0035 | 822.74±0.50 | 1675.8±1.6 | 2.64 | 2.86 | 42.963±0.035 |

To extract precise end-points for the solutions we used simultaneous least square curve fitting of exact analytical expressions for $R$ and $\theta(t)$ derived from spherical Non-Euclidean algebra. We fitted $R$- and Doppler-values simultaneously to expressions for a right-angled triangle with minimum satellite distance $R_{min}$ and flight length $v(t-t_0)$ on a spherical earth for the angle $\widetilde{\theta_6}$ in the lower end of the triangle with 1000 times less weight on the Doppler-part of $\chi^2$ than on the precise $R$-part with free parameters $R_{min}$, $v$ and $t_0$. The analytical expressions in the spherical approximation without satellite movement are given by (Wikipedia-2):

$$\cos\left(\frac{R(t_n)}{R_J}\right) = \cos\left(\frac{R_{min}}{R_J}\right)\cos\left(\frac{v(t_n - t_0)}{R_J}\right) \tag{7}$$

$$\cos(\theta_n) = \frac{\cos\left(\frac{R_{min}}{R_J}\right) - \cos\left(\frac{v(t_n - t_0)}{R_J}\right)\cos\left(\frac{R(t_n)}{R_J}\right)}{\sin\left(\frac{v(t_n - t_0)}{R_J}\right)\sin\left(\frac{R(t_n)}{R_J}\right)} \tag{8}$$

For continuous calculations, $t_n$ is replaced by $t$ and $\theta_n$ by $\theta$. Complete expressions including perturbation from satellite movement are symbolized with a tilde over the variables ($\tilde{R}$, $\tilde{\theta}$). The fitting parameters for the southern and south-eastern routes are listed in table 2 together with $\chi_R^2$ for the $R$-contribution and $\chi^2$ for the entire fit. Exclusively 21:41, 22:41, 23:14 and 00:11 were used for the fits, since only for these points we are sure of a straight flight with constant speed. $\chi^2$ shows the best fit for the south-eastern route. This value is 3.4 times lower than for the southern route (Ashton et al., 2014). The statistical $\chi^2$ due to measurement uncertainty is roughly 1.5 times our best value, meaning both are within the expected uncertainty range, with the southern around the upper statistical limit. The fitted values for $\tilde{R}$ and $\Delta\nu$ are listed in table three where the perturbations are added back facilitating comparison with raw data.

Table 3. Fitted values for $\tilde{R}$ and $\Delta\nu$ where the perturbations are added back, and the Doppler-shift calculated for all relevant points using a southern route for $\Delta\nu$ simplifying comparison with raw data.

| | Solution towards 13.3˚S | | | | Solution towards 34.6˚S | | | |
|---|---|---|---|---|---|---|---|---|
| Time (UTC) | $\Delta\nu$ (Hz) | Deviation (Hz) | $\tilde{R}$ (nmi) | Dev. (nmi) | $\Delta\nu$ (Hz) | Deviation (Hz) | $\tilde{R}$ (nmi) | Dev. (nmi) |
| 21:41:27 | -385 | -18 (±15) [1)] | 1952.96 | 0.46 | -377 | -10 (±7) | 1953.5 | 1.0 |
| 22:41:22 | -522 | -1 (±15) | 2192.77 | -0.43 | -521 | 0 (±7) | 2192.2 | -1.0 |
| 23:14:00 | -594 | 6 (±15) | 2342.3 | - | -589 | 11 (±7) | 2341.6 | - |
| 00:11:00 | -705 | -3 (±15) | 2620.58 | 0.28 | -703 | -1 (±7) | 2621.1 | 0.8 |
| 00:19:29 | -721 | 70 (±15) [*)] | 2664.55 | 2.55 [*)] | -718 | 73 (±7) [*)] | 2664.5 | 2.5 [*)] |

*[1)] Larger deviation perhaps due to passage of a thunderstorm as the plane entered the intertropical convergence zone (Schneider et al., 2014)*

*[*)] Due to flame out (irrelevant for fit quality)*

The last task is projecting the results on earth, which is slightly oblate due to rotation. This is done using formulas for the radii of curvature in the relevant directions and for the relevant latitudes (Wikipedia-1) to convert distances to oblate geometry. Table 4 lists relevant modified $R$-values. After correction, we manually placed the solutions on the earth surface using Google Maps (GM) by demanding that all distances and angles should fit as shown in figure 4 for the south-eastern solution. The southern solution becomes practically identical to the Inmarsat solution with a deviation on the 6th and 7th arcs of only 21 km and an end point at (34.591˚ South, 93.161˚ East). This means that there is no reason to refit the initial part of this solution and fine-adjust the rest, since everything will be practically identical to the findings of Ashton (2014).

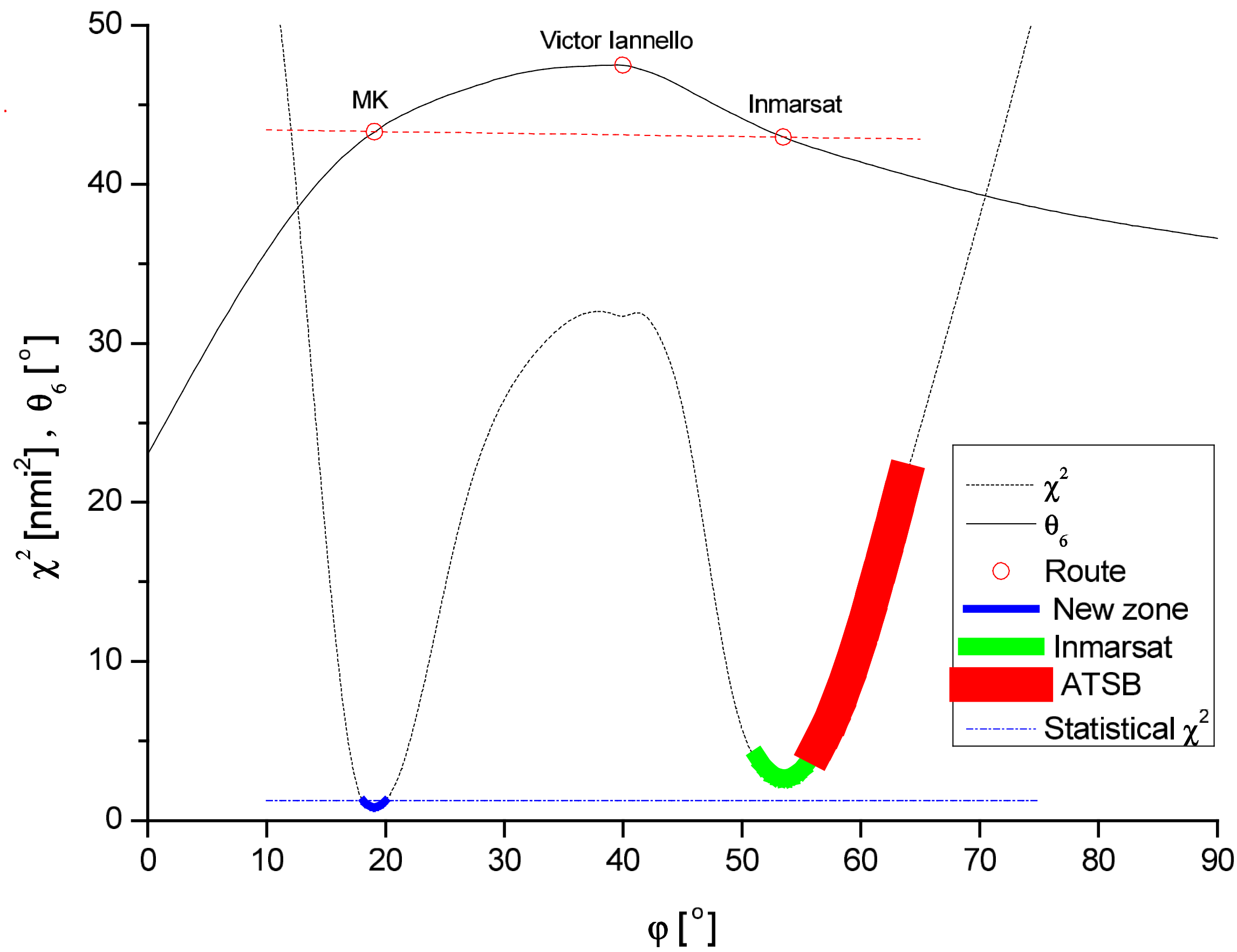

Figure 3. Sketch of the angle $\widetilde{\theta_6}$ (full line) and $\chi^2$ (dashed line) as a function of $\varphi$ (angle from east to south from satellite ground projection at 00:11 to 6th arc crossing) with the search-zones highlighted by coloured lines with thickness roughly proportional to the zone width. Three routes to the south-eastern quarter of the 6th arc are labelled and indicated with red circles, The two stable solutions are connected with a dashed red line to illustrate how the angular dependence determines the local length of the search zone in the same way as dispersion determines the width of an optical filter, where large dispersion gives a narrow filter (Plougmann and Kristensen, 2004). The dash-dotted blue line is the measurement uncertainty. The search zone (ATSB, 2017) had small angular dependence and high $\chi^2$ with small chance of finding the airplane.

Table 4. *R*-values corrected to oblate earth.

| Oblate correction of *R* for placing on earth | $\tilde{R}_{min}^{oblate}$ (nmi) | $\tilde{R}_{6}^{oblate}$ (nmi) | $\tilde{R}_{7}^{oblate}$ (nmi) *) |
|---|---|---|---|
| Solution towards 13.3˚S | 1658.03 | 2618.98 | 2661.6 |
| Solution towards 34.6˚S | 1675.8 | 2612.6 | 2654.6 |

*) *Using average of model and measured value as end-position*

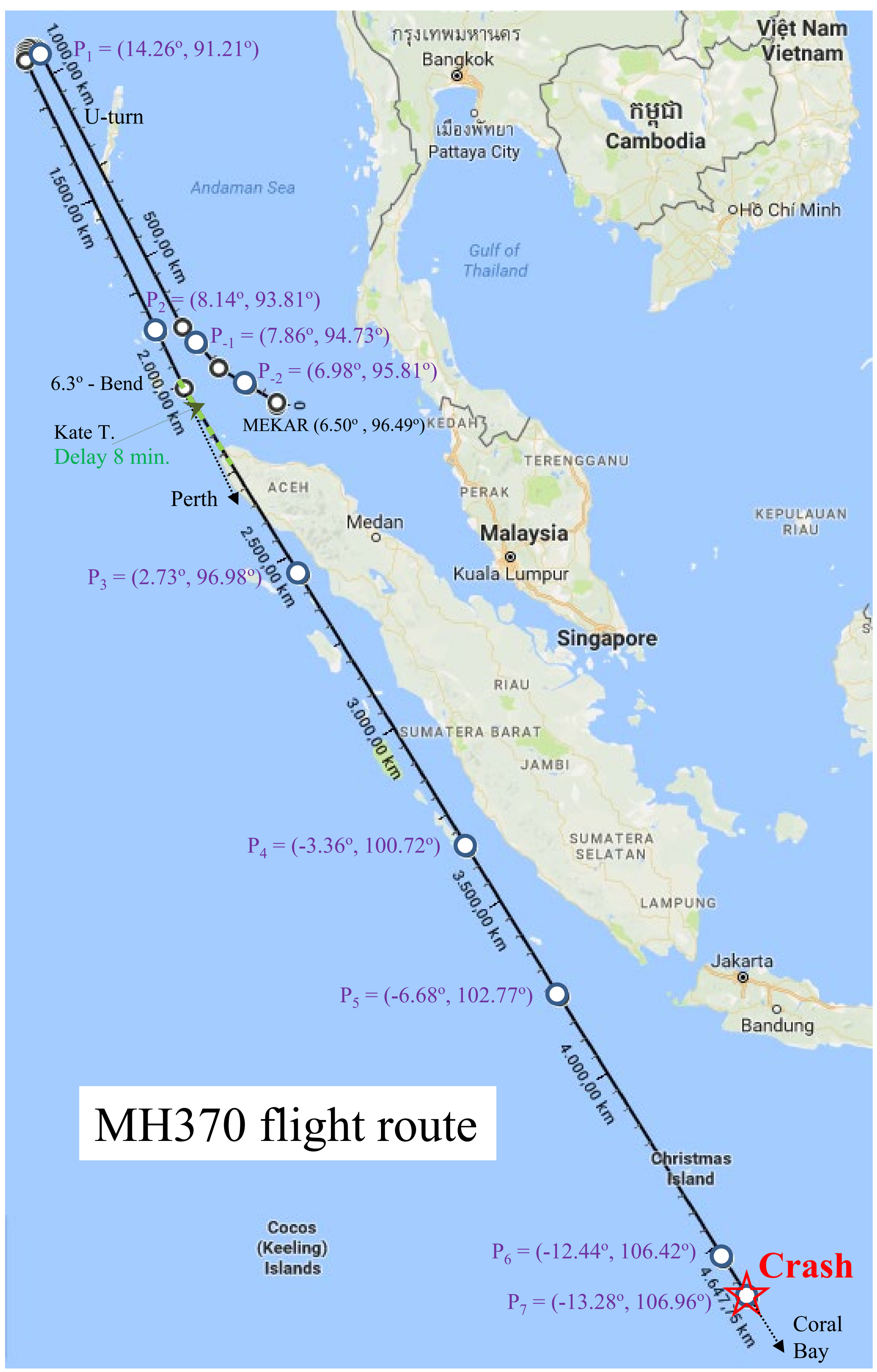

Figure 4. GM illustration of the entire south-eastern solution including the initial part from MEKAR via the U-turn to the merging point near Bandar Aceh with the airplane coordinates when handshakes took place.

For the south-eastern solution we combined the radar data at MEKAR with the handshakes at 18:28, 19:41 and 20:41, and the BFO value at 18:40 with a U-turn soon after 19:41 and a final merger with the straight route soon after 20:41 (with a small delay to match) to construct the initial part of the route. After several attempts, we are convinced there is only little room for different solutions, but minor deviations (up to ±10 km) are possible. In fact, many solutions in south-easterly directions can be made to follow straight lines by introducing a few minutes delay near Bandar Aceh in contrast to the conclusions by Inmarsat, but with significantly worse fitting quality than our particular solution. In addition, we extrapolated the fit to 00:19:29 in the other end and found almost perfect agreement with the *R*-value, while the Doppler-shift deviated significantly downwards, as expected for an engine flameout (Holland, 2017). We choose a middle point as our best guess for the crash-position (13.279˚ South, 106.964˚ East).

3. MATHEMATICAL TESTING OF THE VALIDITY AND PLACING OF THE STABLE SOLUTIONS USING ELLIPTICAL TRIANGLES. We derived analytical solutions to complete an elliptical triangle including satellite perturbations by adding the small triangle formed by the net satellite movement in the upper corner as illustrated in figure 5. The first and last positions are at $R_{min}$ and $\widetilde{R_6}$ (6th arc), where $R_{min}$ is linked up to the most northern satellite position. The small triangle is handled with simple Euclidean geometry. We use the non-Euclidean value for the angular sum in the combined triangle with area, $A$, extracted from GM by back-extrapolation of the $R_{min}$ and $\widetilde{R_6}$ lines to their crossing point $O$ as sketched in figure 5. The resulting equations are

$$90° + \widetilde{\theta_6} + \gamma = 180° \left(1 + \frac{A}{\pi {R_J}^2}\right) \tag{9}$$

$$a_0 = \frac{r_s}{\sin\gamma} \sin(92.679° - \varphi) \tag{10}$$

$$a_6 = \frac{r_s}{\sin\gamma} \sin(87.321° - \gamma + \varphi) \tag{11}$$

where $r_s$ = 115.748 km is the length of the linear satellite projection movement from 19:41 to 00:11, $\gamma$ is the top angle at $O$ for both triangles, $\varphi$ is the angle from east at the satellite projection point at 00:11 in clockwise direction to the airplane position at the 6th arc, $a_0$ is the length of the back-extension of $R_{min}$ to $O$ and $a_6$ is the extension distance of $\widetilde{R_6}$ back to $O$. We again use the tilde to illustrate that the satellite ground projection movement is added back on, since we are here placing the solution on earth using the true satellite projections as fix-points. Equation (9) is based on (Wikipedia-1), while we have derived (10) and (11) using the information above and the data on satellite projection movement. ($\widetilde{\theta_6}$, $A$) assume the values (43.294˚, $6.678 \cdot 10^6$ km$^2$) and (42.963˚, $6.729 \cdot 10^6$ km$^2$) for the south-eastern and southern solutions at the 6th arc respectively.

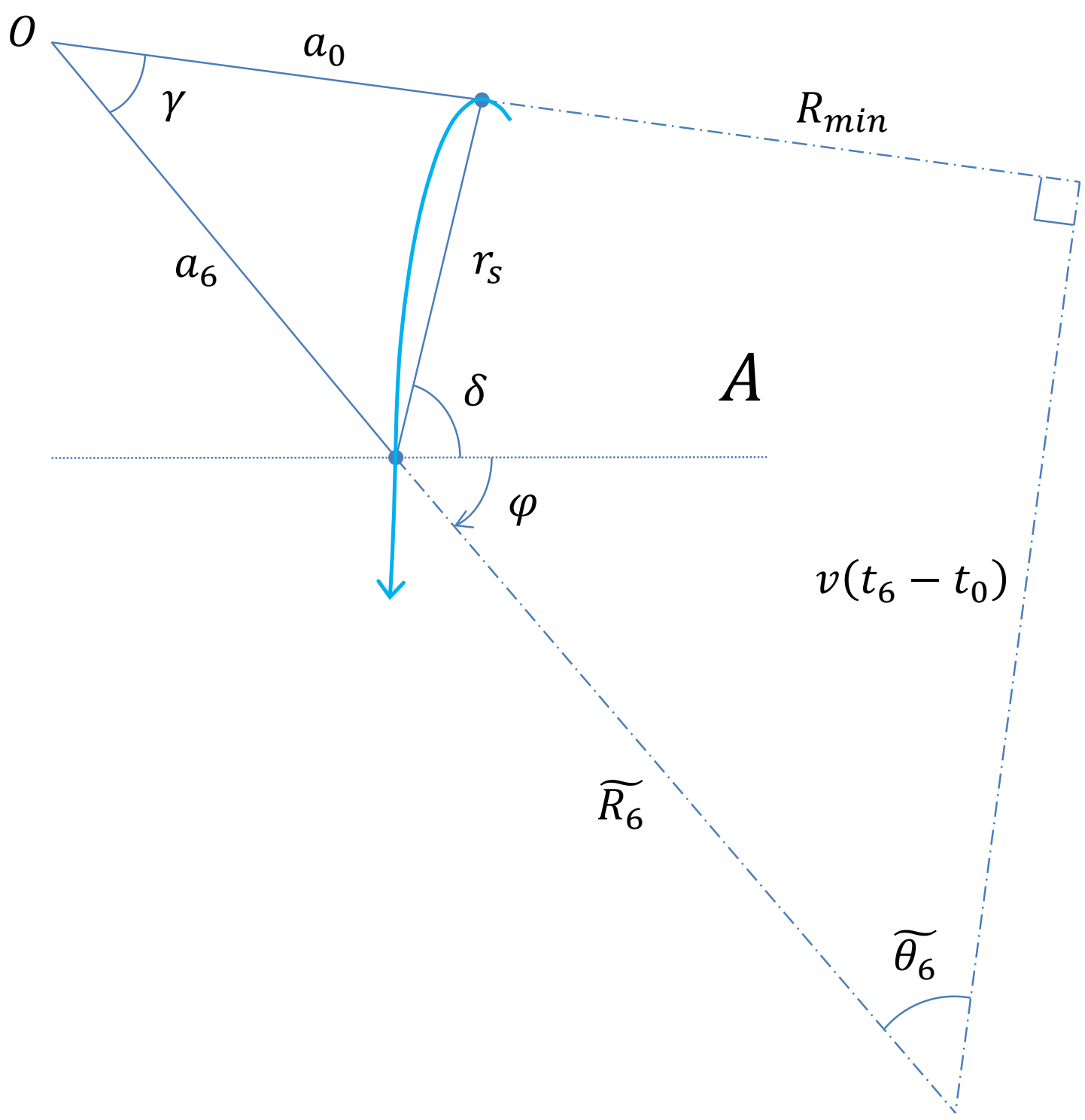

Figure 5. Sketch of the composition of the small (satellite movement) triangle used to complete a spherical triangle for Pythagoras-testing of the placing and validity of the stable solutions. The dash-dotted part of the large right-angled triangle is deliberately drawn too small, and the satellite projection is added as a curved arrow (in movement direction) with the relevant start- and end-points marked by filled blue circles.

Equations (10) and (11) are used as independent checks of the solution (assuming negligible change in $A$ as a function of $\varphi$ near optimum), and the accuracy of the GM placing by comparing the values of $\varphi$, $a_0$ and $a_6$ from GM and (10) and (11). The first is done by insertion into equation (7) which is used as a non-Euclidean analogue of Pythagoras (Wikipedia-2). For the two stable solutions we find $\varphi = 19.3^\circ$ and $\varphi = 51.6^\circ$ respectively by demanding exact validity of (7) for the large triangle. The south-eastern solution agrees within 20 km with GM, while the southern solution deviates almost 200 km. The reason is that this region has large non-Euclidean modification due to rapid changes in $A$ and relatively large oblate correction factors (see table 4). All values of $a_0$ and $a_6$ agree within 0.8 % with (10) and (11), except $a_0$ in the southern solution which deviates almost 6 % (for the same non-Euclidean and oblate reasons). We have hereby performed tests of the validity and placing of our stable solutions.

4. OTHER POSSIBLE SOLUTIONS SATISFYING THE SATELLITE DATA. In order to be completely sure we find all possible solutions (candidate routes) for the model we derive formulas for identifying additional (unstable) solutions. Here larger approximations are made, so the result only allows crudely estimating sixth arc points within 250 km along it. The radial precision is still conserved.

First order perturbation treatment of the average movement of the satellite during the flight tells us that flights directed 3° south (from satellite ground projection) have minimum net sensitivity to satellite movement. All other flight routes will be affected predominantly by the satellite movement perpendicular to this direction. The velocity, $v_0$, going into the Doppler-effect is therefore given by

$$v_0^{(p)} = v + v_{sat}\sin(\varphi - 3°) \quad (12)$$

This velocity will contribute to the Doppler effect through its cosine projection along $R$ given by

$$v_0^{(p)}\cos\left(\widetilde{\theta_6^{(p)}}\right) = v_{Doppler} \quad (13)$$

$v_{sat} \approx$ -50 km/h near 6th arc, meaning that the Doppler effect is reduced for flights in southerly direction, so they need higher $v$ to give the same Doppler effect. The index ($p$) labels three different solution types:

$p$=1: Normal speed routes: The flight joins up with the model-line after the $R_{min}$-point at $t_0$ (so $t_0$ is negative)
$p$=2: High-speed routes: Near airplane performance limit, joining up before $R_{min}$ (giving positive or zero $t_0$).
$p$=3: Low-speed routes: Unrealistic curved routes with small $v$ and $\widetilde{\theta_6^{(3)}}$, typical speeds around 650 km/h.

The optimum values for this simplified model are determined from the three known solutions:

$v_{sat}$ = -56.3 km/h, $v_0^{(1)}$ = 780.5 km/h, $v_{Doppler}$ = 571.2 km/h, $v_0^{(2)}$ = 845.5 km/h

As a test we independently find (due to surplus information) $v_{V.I.}$ = 879.9 km/h, which is close to the value used by (Iannello and Godfrey, 2016). Because the exact starting point is unknown, the largest uncertainty associated with using this simple model is the determination of $\widetilde{\theta_6^{(p)}}$. For eventual extra southern routes, we chose a virtual starting point at a strategic position in the middle of the southern mouth of the Malacca Strait (west of Bandar Aceh) which all routes must (roughly) pass through. Using this we find no additional solutions within the fuel range, confirming that the solution is complete in southerly directions.

For northern directions, we use MEKAR as starting point and find four optional solutions. Two are unphysical (strongly curved and/or zig-zag) routes ending in the Yunnan province of China and one leading to western Kazakhstan is outside the fuel range. The last one towards a 6th arc crossing at (43.87˚ North, 70.06˚ East) in south-eastern Kazakhstan is almost possible, but is ruled out by impossible timing and/or flame-out before 6th arc without unrealistically strong tail wind against the global trade winds (Schneider et al., 2014) during the first half of the route in addition to the difficulties mentioned before.

The solution by Iannello and Godfrey (2016) is timely even though it includes loiter or a U-turn near Perka. However, a large $\chi_R^2$ speaks against it. The total $\chi^2$ is 37.5 times higher than our best solution and 25 times above the measurement uncertainty. It also suffers from relatively poor agreement with some of the additional data. Most importantly, it balances on a mathematical knife-edge. Normally there are two such solutions as in northerly direction but this one lies at the bifurcation point where there is exactly one (double root). A location near the bifurcation leaves little room for improvement. The reason is that it uses maximum speed to match the marginal angle in this area. In conclusion, we find this type of solution an unlikely candidate for the flight route.

After effectively completing this manuscript (Iannello, 2017) published a more elaborate analysis which fits better. However, it displays $\chi_R$ making the deviation look smaller than $\chi_R^2$ and includes all points from 19:41, which lifts the bottom level for $\chi_R$ so the curve looks very flat, surprising some of the bloggers. This illustrates how essential it is to exclude 19:41 to find the south-easterly solution with the U-turn. If one includes 20:41, it is possible to find this solution but the fit is systematically poor without delay and a small direction change between 20:41 and 21:41.

## 5. INCLUSION OF ADDITIONAL DATA AND INFORMATION IN THE ANALYSIS.

We use all other publically available data to choose between the two stable solutions. There are mixed opinions on the debris beaching (Iannello, 2017), but several reports conclude it fits better the further north one gets the crash as long as it is not near the Indonesian coast. This goes for back-tracing of the flaperon from Reunion performed by Geomar in (2015) and (2016), a report from a group of oceanographers (Theguardian, 2016) and back-tracing of temporarily beached debris performed by (Chillit, 2018). In addition nothing was found during the aerial search or the seabed search in the official search zone (including its later extension), which (in combination) effectively covered most high-probability area near the Inmarsat and Iannello solutions. Together these issues make southern routes more than an order of magnitude less probable than the south-eastern route.

Analysis of the flaperon biofouling also delivers important results (Wise, 2018). There was only one species present (tropical Goose barnacles). Furthermore Ca/Mg analysis of a large barnacle shell (Dailymail, 2016) and (ATSB, 2017) shows that it experienced an unusual thermal history with initially very high temperatures then dropping to values near its growth limit of 18˚C and then gradually rising to values typical for Reunion. No one has been able to come up with a good explanation for this peculiar result. However, looking at sea currents and temperature world maps (Hunter, 2013) this kind of behaviour is possible when starting out near our south-eastern solution in the fall. Here a weak current of hot, nutrient-poor tropical water from north carries water towards south amplified by hurricanes (Chillit, 2018) and global warming (Feng et al., 2013), where it meets and mixes with the relatively cool and strong Western Australian current coming from the south. The mixed current continues via Reunion to Africa under heating by tropical sunshine. Therefore, debris coming from this area is settled mostly by tropical barnacles and will experience such a temperature profile when starting in March. The temperature drop is intensified and prolonged by the onset of winter. Therefore, the barnacle results add roughly another order of magnitude preference for the south-eastern solution.

The only weak point of this simplified analysis is that the barnacles seem to be much too young (only a few months) judging from their size. However, one must remember that their growth-rate depends on two parameters – temperature and available nutrients. During the initial part of their life the temperature was ideal for growth, but with only small amounts of nutrients leading to a relatively slow growth. During the middle part, they practically went into hibernation due to temperatures approaching their growth limit. Finally, they drifted towards Reunion and grew increasingly rapidly. This explanation also clarifies another issue, namely why some of the barnacles grew above the water line. Most likely, there were temporarily a much larger number of tropical barnacles on the flaperon weighing it down and preventing settlement of other barnacles during the winter. However, during the cooler period most of the initial barnacles died and later fell off leaving a few alive above the average water line. Most likely the same happened for other debris, which explains why it was much cleaner than usual, leading some investigators to believe that the debris was planted. In case the crash had happened at the Inmarsat position, a variety of different biofouling with origin in increasingly warmer climate-zones would gradually have covered the flaperon, resulting in much higher biodiversity.

However, the line of circumstantial evidence does not stop here where some readers may already be convinced. Depending on the finer details the official investigators estimated (Holland, 2017) that the crash starts between 00:19:29 and 00:19:37 where the plane is losing height somewhere between 15700 feet/min (fpm) and 25300 fpm indicating insignificant pilot control. The average downward speed will most likely be below the middle of this interval since a crash at 25000 fpm would probably give smaller pieces of debris than observed (Wise, 2018). In any case an uncontrolled crash gives rise to a random horizontal walk near the end-point of the line reaching the sea surface 1-2 minutes later (Steel, 2016). This is in excellent agreement with a sound feature recorded at the nuclear arms listening device HA01 near Cape Leeuwin (34.892˚ South, 114.153˚ East) at 00:49:42. With a water-sound-speed of 1.484 km/s expected along this passage (Steel, 2016) the plane should crash 114 s after 00:19:37 at our south-eastern solution to match the recording. This corresponds to a downwards speed of roughly 15000 fpm and a crash time of 1:54 min. New contrail data point to a crash 20-40 s later. A moderately rapid crash fits the average fragment size. In contrast, no sound features agree with the southern routes.

Finally, the eye witness (Tee, 2014) describes a large airplane diving to low height and flying slowly west of her boat located north of Bandar Aceh at (6.628˚ North, 94.438˚ East, plus 15 km east-northeast due to later time) on the night MH370 disappeared. It came from north, made a moderate turn nearby towards her and disappeared somewhere south without landing. The plane had a red halo around it and the normal lights and windows could only be seen in the cockpit while the rest looked strange. Considering diffraction of red warning light and small windows in the cabin versus white (green) light in large cockpit windows puts the diagonal distance between her and the plane around 2.5 km. She estimates a 3-km horizontal distance. These observations agree with our south-eastern solution, where the plane makes a 6.3˚ turn 17 km north of her position while diving and causing delay, and passes 2 km west of her boat. Only the time does not fit. She puts the closest approach at 19:20 while our model says 20:59. This is also essential to get spatial agreement due to the movement of her boat perpendicular to the predicted flight route. However, she was particularly uncertain about the time, so it is not an unlikely error in an area where local time-zones in India and Indonesia deviate two hours within a few km of her position.

## 6. DISCUSSION.

For the optimum Christmas Island solution, we estimate that the model-placing-uncertainty is ±35 km along the sixth arc. The maximum error will therefore be roughly ±70 km ($2\sigma$), which we use as the (half)-length of the search zone. For the transverse extend we propose ±15 km, since the largest contribution to this uncertainty is second-order placing error (found experimentally to be 1/6 leading to ±6 km) followed by random walk uncertainty after 00:19:37 (±3 km) and fitting uncertainty (±3 km) giving a total of ±7.5 km transverse uncertainty, and again choosing the double for the extend of the search zone. It is worth noting that all data point to positions within the central 10 % of the search zone with the largest deviation coming from Kate's observation leading to a point 4-8 km south of the centre of the zone. As a funny coincidence a 5-km shift south will perfectly align the straight flight with an end-point at Coral Bay Airport and roughly remove the 8-km offset at MEKAR (Iannello, 2017). We therefore guess there will be a relatively high chance of finding the plane within 350 km$^2$ out of the 3500 km$^2$.

There is also something special about the Christmas Island route going through the intertropical convergence zone (Schneider et al., 2014) where satellite detection and long-range radar are hampered by tropical thunderstorms, indicating intelligent planning. Most likely, the perpetrator(s) also knew about the handshakes and deliberately directed and timed the flight to get close to the worst possible mathematical data-entanglement with satellite movement through spatial correlation, making it almost impossible to find the plane because this allows for a multitude of solutions with similar fitting quality. This was achieved by flying a route resembling a magnification of the satellite ground projection curve with the SDU restarting as the plane left Malaysian radar coverage and entered this route. The U-turn was carefully aligned to match the top of the satellite projection curve, and the immediate continuation was slightly curved, followed by a straight flight to the end, pointing to perpetrator(s) with knowledge of entanglement (Wikipedia-3) and (Kristensen et al., 2012). Interestingly, flight simulation data found near the captain's private computer (Steel, 2016), (Iannello, 2017) and (Wise, 2018) resembles a classical analogue of a quantum Singlet-entanglement (anti-correlation) with the satellite motion while the actual route matches a Triplet-entanglement (correlation) providing optimum hiding (Kristensen et al., 2012). However, combination of several scientific methods with topology optimization (Plougmann and Kristensen, 2004) and (Bendsøe and Sigmund, 2003) allowed discovery of the Christmas Island route and its identification as the best solution.

In case one would also go for a repeated search near the end of the southern route, the placing-uncertainty is 2.5 times larger due to non-Euclidean effects. This leads to a search zone area of 20000 km$^2$ which is close to the original official estimate of 25000 km$^2$ (ABC, 2017). However, most of it was already part of previous search zones so a much smaller area will be sufficient.

To further strengthen evidence for our solution it is possible to do one of the following things:

1) Ask how the received signal strength can go up with increasing atmospheric travel from 20:41 to 00:11 with reference to (Inmarsat, 2015) coverage and an antenna model?
2) Look for coincidence with sound recorded at Scott Reef or HA08 near Diego Garcia. Unfortunately, data for the relevant times is not publically available (and Diego Garcia data compromised by local noise from a military exercise as shown after our publication).

If we find one such coincidence, classical triangulation pinpoints the exact crash-site to a few km. However, it is the local in-coupling in the sound-guiding layer of the ocean, which is most important – not the distance. Underwater mountains north of the expected crash-site may cause most coupling in southern direction and add some confusing echo. For a weak signal, this potentially prevents identification at Scott Reef.

After effectively completing this manuscript, a method was published for analysis of sound propagation in water to determine impact-distances with only one detector (Kadri et al., 2017) including application for a partial re-analysis of the data from HA01 on the night MH370 disappeared. There is a deviation of 2 minutes between their time axis and previously published data from HA01 (Steel, 2016). If the new axis is correct, MH370 crashed exactly as the last handshake was interrupted rather than 114 s later, and the distance to the satellite projection would be roughly 11 km shorter due to smaller $h$, predominantly via equation (3). However we suspect there is an error in (Kadri et al., 2017) since the distance calculated for the stronger neighbouring peak seems to be off by roughly 200 km as pointed out by some bloggers (Iannello, 2017) consistent with a 2-minute error. Alternatively it is possible that the two signals accidentally coincided, which may explain the distance-discrepancy by interference.

7. CONCLUSION. In conclusion, we have used a novel combination of methods from science and engineering to disentangle and discuss all four solutions to the model of the disappearance of flight MH370. For the two stable solutions, we have delivered rigorous proof of their placing and validity by using a non-Euclidean version of Pythagoras. All other publically available data point to the Christmas Island solution, while we rule out the other three. The southern route is second best but still unlikely, and the decision to stop searching was correct (ATSB, 2017). Resumed searching at that location makes only little sense, since the probability of finding the airplane is below 1 % according to our estimates. We propose instead a new, focused search zone of 3500 $km^2$ centred at (13.279˚ South, 106.964˚ East) with slightly elliptical shape along the seventh arc and a total length of 140 km and width of 30 km. The probability of finding the plane there is above 90%.

After completing the manuscript, we found additional evidence for the Christmas Island solution in 2019. In 2024, we found black shadow contrails and a mushroom-cloud from the crash as described below. This shifts the crash position to (**13.53° South, 107.11° East**), reduces the search area to a 10-km radius, and increases the probability to 99%.

8. NOTE ADDED IN PROOF IV. Based on discussions with several bloggers (Iannello, 2018-2) and other independent investigators after the publication of the first, second and third versions of our manuscript on 'How to find MH370?', and additional research into specific details, we decided to rewrite the 'Note added in proof' paragraph to update the information a third time in 2024, including an appendix on contrails. The rest of the original manuscript remains practically unchanged with a few minor updates and corrections.

Initially we reanalyzed the connection between different routes and the observations by Kate Tee using the speeds found by Victor Iannello for different end-positions at the seventh arc (Iannello, 2017). Hereby we found that only routes ending between 26° S and 32° S could be consistent with her observations if they took place between the particular gybes she pointed out. Only routes leading to latitudes between 11° S and 16° S would be consistent, if her observations took place after the last gybe. All other possibilities lead to inconsistency with her observations.

As part of the same analysis, we looked at possible agreement with the sound-features at Cape Leeuwin. The most likely feature is only consistent with end positions between 12° S and 15° S, while one of other two (much less likely) peaks could move the end-position down to 16.5° S (maximum). All other proposed end-positions are inconsistent with the observed peaks. Trusting the input from Kate Tee (except the before/after gybe issue) this makes our solution at 13.53° S most likely, and practically rules out any solution between 20° S and 25° S. A statistical analysis including information from e.g. the unsuccessful seabed searches (details not included here) leaves only 2 % probability of finding the wreckage between 20° S and 25° S. New contrail data from 2024 reduce this to 0 %.

Secondly, it became clear from discussions with bloggers (Iannello, 2018-2) that the method we presented for calculating the BFO for different routes than the Inmarsat solution was inaccurate. It needed an additional correction for the local speed of the airplane. Furthermore, it is unclear if the method we used for handling a pressure-induced shift (presented in the first version of this paragraph) is correct. Concerning both these issues, it is extremely important to keep in mind that the errors are relatively small (in particular the pressure shift), and that due to the small weight put on the BFO in our topology-optimization, these errors will only lead to small shifts of the final position. However, it is still important to make the corrections properly for two reasons. First of all, there is a risk that the effects may exceed the perturbative regime, so another one of the four possible end positions could overtake the role as the best fit. Secondly, even if the corrections only lead to shifts within the measurement uncertainty, they may still add significant extra search time and cost for a resumed search if handled incorrectly.

The procedure for the speed correction is simple to first order. Because the speed on the route towards 13.53° S is 3.1 % lower than for the Inmarsat solution, the net (internally calculated) Doppler shift compensations reduce by this amount. The first-order corrections for this bring our BFO values very close to those calculated by Victor Iannello (Iannello, 2018-2). There is still an insignificant deviation around 2 Hz due to smaller effects. In the following, we will ignore this since it is far below the statistical uncertainty. However, the 3.1 % correction is clearly significant, and in case one only looks at the BFO (letting the weight-factor on BTO go to zero), one finds an optimum fit somewhere between Victor Iannello's original 27° S solution from 2016 and the most recent independent group recommendation of a position near 34.4° S (Iannello, 2019).

Since the BFO values are much less accurate than the BTO values, and their detailed interpretation is somewhat uncertain, this is clearly not a good approach. By instead choosing to keep our original 1000-fold lower weight on the BFO, one gets a much smaller shift (around 20 km south) near our original 13.3° S solution (fitting 13.53° S). This seems like a completely insignificant shift, but it is unfortunately still so large that it becomes marginal to use a first-order perturbative approach, and the chi-squared values for the two best solutions get closer to each other. It is unclear how to handle this situation without a complete reanalysis from the bottom, but before doing this, it is essential to consider the impact of other perturbations from changes in pressure and temperature in the cabin.

Looking at the 13.53° S solution it is obvious that the entire solution only makes sense if an attempt to bail out by parachute took place while flying slowly and at low height near Bandar Aceh. This would have left one of the doors behind the wings open, leading to significant drops in pressure and temperature as the airplane returned to 11 km flying height. It is unclear how to handle the pressure change, but since it is most likely around 2 Hz (average value), we choose to ignore it using the same arguments as for the 2 Hz deviation above. However, the temperature change is clearly significant. Bloggers (Iannello, 2019) found a shift for the temperature-stabilized oscillator of 0.3 Hz/K. In addition, there is an expected shift in the power-output for the SDU of -0.05 dB/K.

One of the bloggers at (Iannello, 2019) calculated a temperature drop of 60 K inside an open cabin. We largely agree on this number provided the heat supply systems are 'off'. However, one also needs to know the detailed temperature curve. We found this by assuming a linear cooling rate and a thermal time constant $K^{-1}$ around one hour. The exact value of $K$ was determined by solving the thermal differential equation

$$\frac{\partial T}{\partial t} = -K(T - T_{11km})$$

and fitting to the result, where $T_0$ = 25°C and $T_{11km}$ = -45°C, giving $K$ = 0.87 $hour^{-1}$ from

$$T = T_0 - 70°\left(1 - e^{-K(t-t_0)}\right)$$

If the cabin is cooling for 3 hours with a one-hour thermal time constant, this gives around 65°C temperature drop. Below we describe how to find the exact value of $K$.

In order to perform the thermal fitting we used the additional (and practically overlooked) parameter measured by the Inmarsat system, namely the received power. In order to use this, it is necessary to develop a model for the power transmission to 3F1. We developed such a model based on a 3-step procedure. We estimate the diffraction loss using the Fraunhofer approximation. We use angular momentum projection to estimate the circular polarization overlap, and we estimate the effective antenna area using geometric projection.

In an earlier first-order attempt, we used an emission pattern calculated by (Harvey, 1963) under the assumption that the antenna was of the oldest (static) design type as described in the paper by (Fu, 2012). After communicating with bloggers at (Iannello, 2019) we were informed that the airplane used a mechanically adjustable phased array antenna for pointing the beam. We therefore developed a model for this antenna type using the method described above, while now ignoring pointing errors. This model fits much better to the data than the old static antenna model, confirming the validity of the information. By comparing to the received power at the previous flight between Beijing and Kuala Lumpur, we observed an additional 1-dB deviation from front-back mirror symmetry (highest output in forward direction), which seems reasonable given the detailed geometry visible in pictures of this antenna type. We fitted the complete model to the last two handshakes before the airplane disappeared, the two points at 19:41 and 20:41 and the four middle points from the previous flight (Beijing to Kuala Lumpur), meaning 8 points in total, with the power-offset as the only free parameter. For fitting of the previous flight, we used flight parameters from (Davey, 2016) and exclusively values measured in channel 4. However, we ignored points measured very close to take-off or landing (where the exact climbing/decent and tilt angles are important but unknown), and the points directly after reboot (where the temperature is unknown). All 8 points fit within 0.2dB, which is nice considering that the expected uncertainty is around 0.3dB. It is important to notice that the model corroborates that the airplane was approaching 3F1 at 19:41, since this point is only 0.1 dB off from the model when the 1-dB direction asymmetry is included (if not, it is 3 sigma off, - equal to 0.9 dB).

We used the antenna model alone to determine the expected power for all the following points after 21:00 and found poor agreement. However, by including the thermal model with roughly one hour time constant the agreement dramatically improved. We completed the work by allowing the thermal constant to vary freely and found the best fit with $K = 0.87\ hour^{-1}$. We have plotted the results in Figure 6.

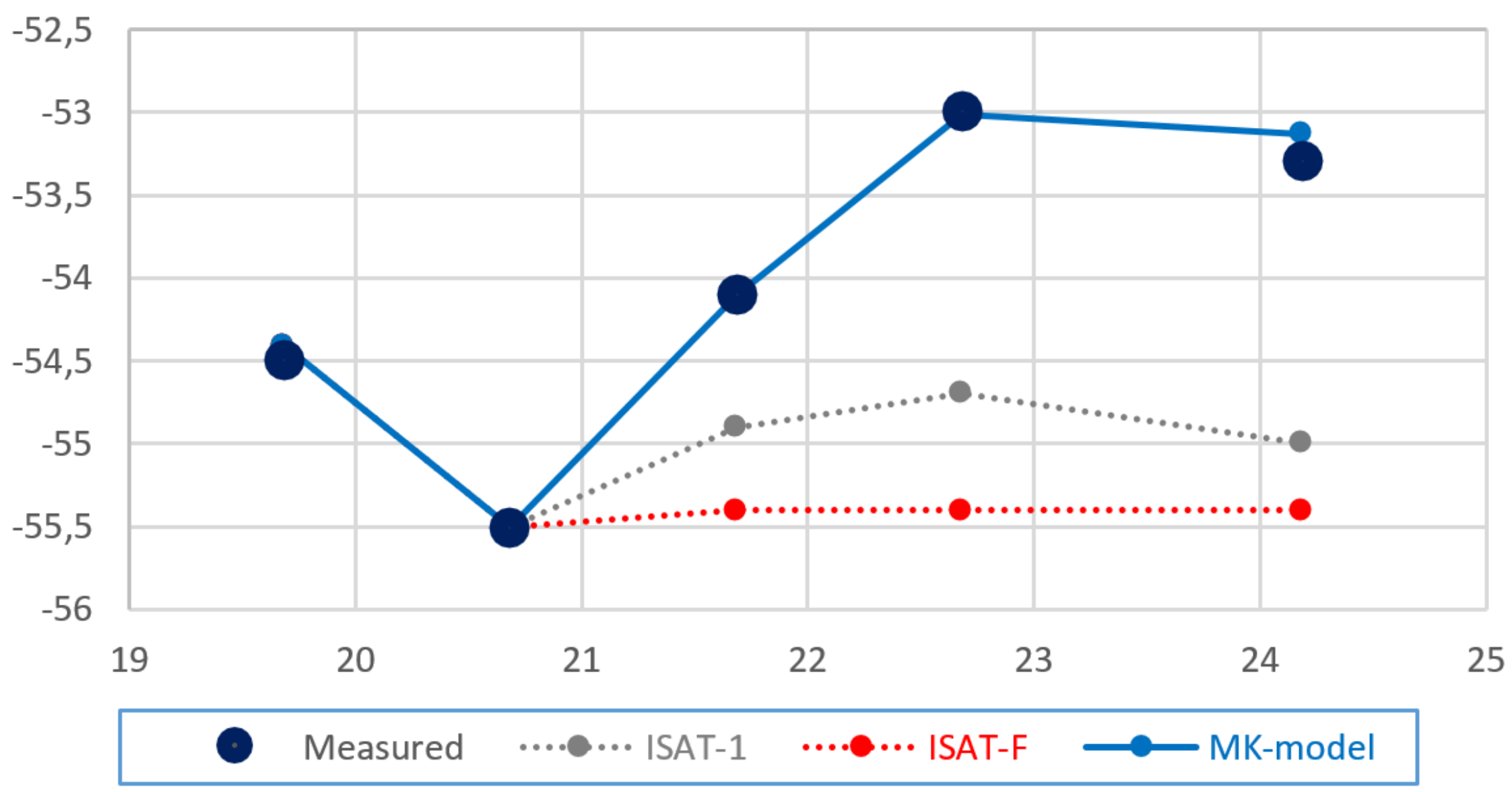

Figure 6. Illustration of the agreement between measured power at 3F1 (large black circles) and predicted power for three different cases. The (small) blue circles connected with blue line is our model including the antenna transmission and cabin cooling due to an open door leading to increased emission power (-0.05 dB/°). The last point is corrected in accordance with figure 8. The grey curve is the expected power assuming a curved flight as in the first estimate by INMARSAT (highest probability version). The red curve is the power expected along the final INMARSAT route, F (and/or the slightly shifted proposal from (Iannello, 2019)).

Maximum deviations are 2.5 dB (8 sigma) for the red curve and 1.5 dB (5 sigma) for the grey curve. In principle, the grey curve improves by shifting it further north, but this quickly pushes the flying speed below the stall limit and makes such a route impossible to follow. The red curve has unacceptably poor agreement with the measured values, while the agreement with the blue curve is practically perfect.

Furthermore, some of the bloggers at (Iannello, 2019) criticized our model for having too large BFO deviations. While these deviations may be due to unusual tail-heavy statistical behavior of the SDU oscillator after exposure to low pressure and temperature between 17:07 and 18:22, it may also be due to cooling after 21:00. In order to test this we calculated the expected oscillator shifts (0.3 Hz/°) for the SDU during the cooling found to match the received power levels in Figure 6. We used the systematic (3.1%) BFO shifts from our model results and plotted these two data sets against each other in Figure 7.

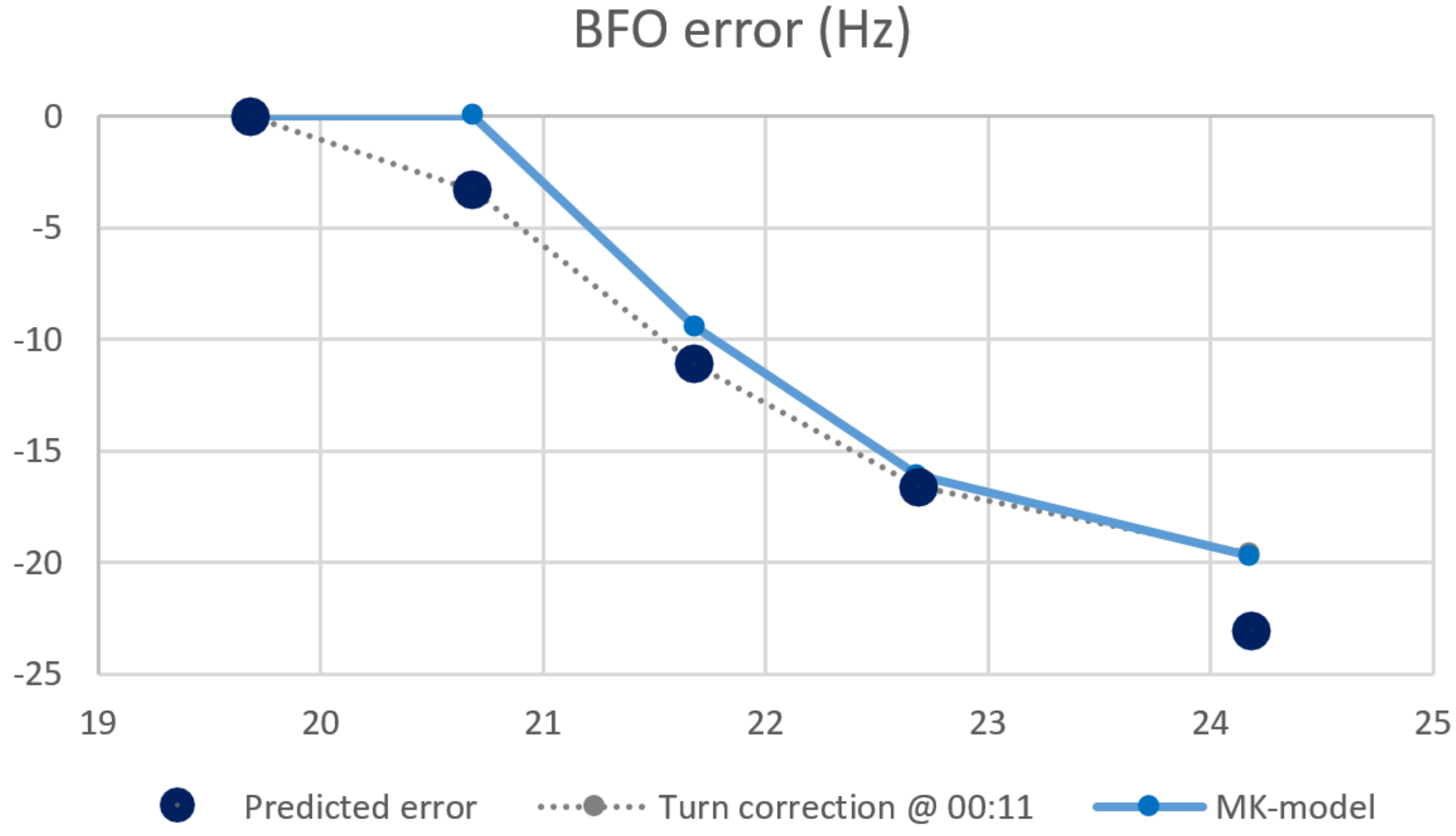

Figure 7. Illustration of the agreement between systematic (predicted) BFO errors and results deduced from our thermal model with an open cabin door. The last point is corrected in accordance with figure 8 (grey). Notice the smooth decay of the small error indicating that the remaining deviation is potentially due to a memory effect from the previous temperature and pressure cycling. It is also important to notice that this figure ignores the random deviations from table 3. The largest one of these @21:41 is most likely due to passage through a tropical thunderstorm in the intertropical convergence zone.

The plot shows good agreement between our thermal model and the systematic BFO errors. This corroborates that the deviations are most likely due to a dramatic cooling of the cabin from an open door. The only larger systematic deviation (4 Hz) occurs at 20:41. This may either be due to the tail-heavy drifting effect or indicate that the cabin door opened just before 20:41 while the airplane was near 2-km flying height (leading to a small initial cooling). We find the drifting explanation most likely, since the deviation slowly decays to zero over the following 3.5 hours, while the power output shows no such behavior (as expected, since the power amplifier has no memory effect).

Finally, we use the calculated temperature drop to estimate the average oscillator shift for all points after 21:00 to -15.5 Hz. This means that we must subtract 15.5 Hz from the calculated (negative) Doppler shift to correct for the average effect. As part of the development of the original version of this paper, we derived a simple formula for the end-position shift due to a 1 Hz shift for all the last four points. This reads one-degree latitude (towards north) to compensate each negative Hz added. The Inmarsat paper comes out with a 25% larger value, but that is for one point individually, and the continuity and straightness of the curve reduces the collective value by roughly 25%. This means that any solution calculated based on BFO (alone) must be shifted 15.5 degrees (latitude) north to correct the effect of the temperature drop. Therefore, the optimum for a purely BFO based solution without temperature drop would be around 29° S, which is not far from proposals from the independent group (Iannello, 2019). The remaining difference could be due to drifting, random errors or low pressure.

At 00:11 the first engine had flamed out, so the airplane automatically turned attempting to compensate the (maximum skew) engine push. A simulation at (Iannello, 2019) indicates a flame out for the first engine around 00:08. We used this to correct the microwave power loss for the last point at 00:11. New 2024-data gives only small additional corrections.

Now the big question is if this is a wild speculation or a fact. First, we looked at the reboot at 18:25. A couple of experts (in particular the blogger DennisW at Iannello's blog) had been complaining vigorously about the normal booting for the SDU. If the left power generator had been off to allow faster flight at high altitude along the Thai-Malaysian boarder, the SDU oscillator would have been extremely cold (since its heater was also off) and at low pressure during the SDU booting, which would lead to severe oscillator drift for a prolonged period. Never the less the booting is normal as shown by Holland. A simple explanation is if the SDU was re-connected at a later (convenient) time when the cabin was back in normal condition but now manually to one of the power supplies. This is consistent with Radar observations between Penang and MEKAR of an increasingly normal flight along waypoints after the fast, high, and slightly noisy (manual?) flight before Penang.

While this is in good agreement with the SDU connected back to a generator, there are still some remaining details to settle. The described succession of events leads to a clockwise crash spiral similar to case 4 in the list of Boeing simulations if the SDU is connected to the left generator (in contrast to most of the Boeing simulations ending in counterclockwise spirals), and places the debris roughly 13 km further south and somewhat west. We initially guessed 10 km west, partly from arc curvature, partly due to height loss at 00:19. In combination with our previous estimate of a 5-km general position error, this gives an 18-km south and roughly 10-km west shift compared to our original position if we ignore the turn in direction before 00:11. Including this turn and a longer than expected straight section adds roughly 18-km additional shift towards west giving a total of 28 km. However, there would be no normal handshake at 00:11 with the SDU connected to the right generator, and the straight flight segment after 00:08 would be too short to fit the second engine flameout with too little remaining fuel.

However, some experts still feel this entire explanation is somewhat speculative. In order to deliver a definitive proof, we therefore used our refined knowledge of the end-scenario to look for matching contrails from the airplane in satellite pictures. New evidence from 2024 indicates that the crash fits better with the majority of the Boeing simulations.

We found several aligned contrail segments matching our solution until briefly after 00:00 in two consecutive pictures from the METEOSAT-7 weather satellite (Weather Graphics, 2014), followed by turning at 00:08, and ending in a spiral. Figure 8 illustrates this contrail with the interesting added feature that it becomes abruptly thinner and more intense at 00:08, exactly as the right engine flames out, the power of the left engine increases to maximum and the turn takes place.

The satellite pictures also contain a contrail from another flight coming from southeast and flying directly over the airport at Christmas Island. Based on the measured speed, timing and angle (extracted from contrails in consecutive pictures) we have identified this contrail is either coming from a flight leaving Melbourne at 5:00 am local time heading for Dubai (EK409) operated with an Airbus A380 explaining the more intense contrail than the one form MH370, or alternatively a cargo flight from New Zeeland. A cargo flight is most likely also using a large airplane type with similar intense contrail.

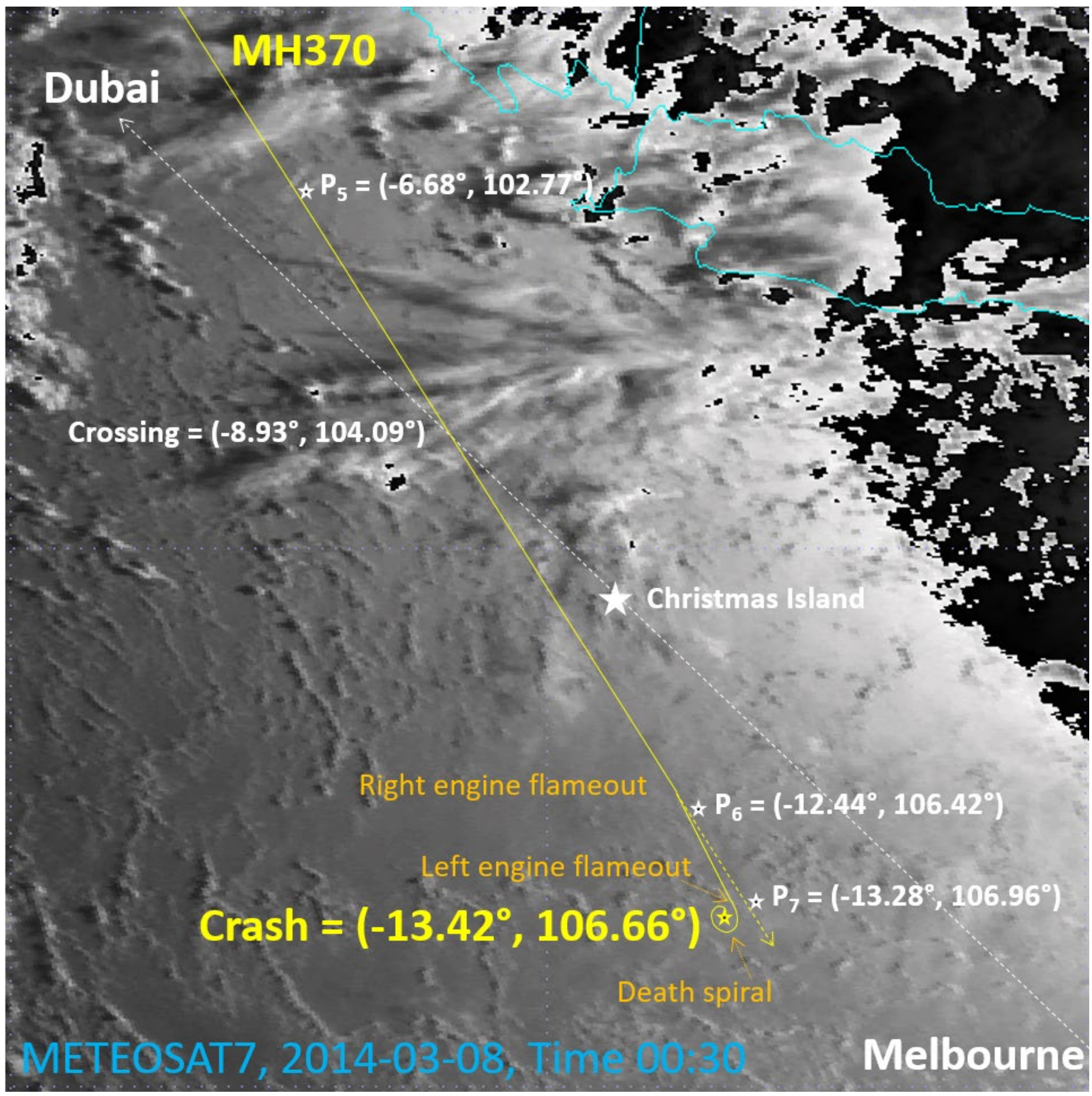

Figure 8. Illustration of two contrails near Christmas Island shortly before MH370 crashed in the region. The dashed white line may be flight EK409 from Melbourne to Dubai or a cargo flight from New Zeeland. The orange line represents a flight coming from north-northwest, making a moderate turn at 00:08, and ending in a spiral south-southeast of Christmas Island. The times and positions of this flight agree roughly with our model (white stars 5-7) for MH370. The turn time agrees with predicted flameout for the first engine within 19 seconds (Iannello, 2019), and the simultaneous reduction in thickness and increase in strength agree with one of the Boeing simulations. In 2024, we found new evidence using shadow contrails as described in the appendix 'Black Swan'. The absence of a shadow contrail for the last part of the route above (after 00:08) indicates this segment was a false contrail, most likely induced by transformation of the satellite pictures to rectangular coordinates. Instead, there is another contrail-segment connected to the upper part of the route near Christmas Island (completely unchanged) with corresponding shadow leading to a position 16 km on the other side of the 7th arc (after turning left instead of right @ 00:08). The new crash position is at **(13.53° S, 107.11° E)** 375 km from Christmas Island, where the shadow contrail merges with its corresponding normal part, followed by the appearance of a white mushroom-cloud under the dash-arrow a few km southwest, and remaining visible for more than one hour. The 'Death spiral' is qualitatively similar to the one above, but tighter and shifted 50 km east-southeast after passing nicely through $P_6$ and $P_7$.

The end position in the middle of the spiral is marked with a yellow star. The scenario leading to crash is similar in the 2024 update (concerning spiral orientation), but the turn at 00:08 goes east (left), the spiral is somewhat tighter and it is moved 50 km east-southeast.

Two seismic detectors (XMI, 2024) confirm both contrails passing Christmas Island in Figure 8. The airplane from the south passing directly over the CI airport gave large signals at the expected time in the seismic detector near the main building of the airport. We did no detailed analysis of this data, but concluded from the low frequency that it was a large, heavy airplane (bigger than MH370) flying somewhat faster than MH370.

Both seismic detectors recorded the passage of the airplane from north-northwest, providing a lot of detailed information. The center frequency (around 12 Hz) is typical for airplanes near the size and weight of MH370. The exact signal arrival time from its shortest approach to Christmas Island fits our model to better than 2 seconds precision (completely outstanding). The Doppler curve from the nearest detector at a low-noise, isolated location in the southern part of the island fits with the speed to better than 4 %. The sign of the internal delay between the two detections prove that it was on the correct side of Christmas Island. The signal of the detector at the airport had its first part chopped off by a hill near the airport, proving that the airplane came from the north. Finally, correlation between the two Doppler curves and the observed contrail confirm the distance of the flight to better than 6 % (3 km) and the angle to better than 10° agreement with our model. This entire collection of evidence from the seismic signals prove without any significant doubt that the airplane must have been MH370, since airplanes rarely pass on that side of the island.

In hindsight, this detailed solution also solves a number of smaller contradictions in the data. First, the downward acceleration found by the Holland paper around 00:19:37 seemed inappropriate compared to the Boeing simulations at that stage of a crash. Knowing from the power level at 00:11 and the contrails that the first engine flamed out a few minutes earlier than initially expected (around 00:08) brings this in better agreement with the Boeing simulations. Secondly, the straight flight from 00:08 to roughly 00:19 is longer than the Boeing simulation. We think this is because of energy saving for the second engine since the pressurization and heating systems were switched off shortly before 21:00. Thirdly, the termination in a tight spiral dive after the contrail-end rules out an active pilot ditching the airplane far from the seventh arc. Fourth, the 15.5 Hz temperature-induced (average) shift partly explains why the initial BFO plots from the Inmarsat paper deviated 10-15 Hz from the measured values. Finally, the crash after a long flight to fuel exhaustion practically proves that it was a deliberate act, and not a technical accident.

As an added curiosity, the contrail from MH370 crosses the northbound contrail at an angle of 13 degrees roughly 240 km northwest of Christmas Island, and the two airplanes might have been close enough for visual contact in light from the rising sun. It would be highly interesting having a chat with the pilots of the other flight, since they might have seen MH370.

Concerning agreement with the peak recorded at Cape Leeuwin, the appearance of the mushroom-cloud indicates the crash happened 20-40 seconds later than initially though. It also happened closer to Cape Leeuwin moving things 20-25 seconds the opposite way. The remaining net shift towards later time (7.5±11 seconds) is unlikely to have significant impact, so the agreement remains good within the uncertainty.

Concerning the beaching pattern of debris, the westward shift due to prevailing currents after the crash would bring the debris close to the maximum wind-field of hurricane Gillian two weeks later. We are unable to calculate the details precisely, but qualitatively it will improve agreement with observations due to some debris trapping in the hurricane, and thereby dragging several pieces to near 20° S where the hurricane decays.

A few open questions remain. Most importantly, whether the perpetrator(s) knew leaving a door open shifted the BFO so it roughly matched the simulation in captain Shahs computer, and knew to eliminate suspicious oscillator drift and inconvenient handshake timing by turning the SDU power supply back on at an optimum time. If this were the case, we are dealing with a highly sophisticated, well prepared, and deliberate act of mass killing with scientific input to manipulate the data and mislead the investigators, not a simple attempt to hide a suicide or an act of terrorism. The motive and fate of the perpetrator(s) remain open questions. However, the nice agreement between our thermal modelling and the measured microwave power levels practically proves that he/they attempted to bail out by parachute near Bandar Aceh. The thermal modelling also explains the unusual statistical distribution of BFO values recorded during the second attempted telephone call at 23:14. The distribution looks like the upper half of a Gaussian, indicating that low temperature and pressure far from normal operating conditions affected the SDU frequency reference.

One final piece of circumstantial evidence also points toward a carefully planned act. The exact crash position is almost perfect to maximize the time for debris to reach flight- or shipping-routes. Looking at an integrated plot of flights recorded by Flightradar24 across the entire world, we find that the first debris will cross flight- or shipping-routes more than two months later, and most of it would beach in Somalia (Iannello, 2019) in the absence of hurricanes, and never become available for the investigators. The passage of hurricane Gillian further masked the crash position and helped mislead the investigation, since it made the beaching pattern look as if the debris originally came from around 20° S. The Flightradar24 data also confirm few flights pass on the relevant side of Christmas Island.

9. APPENDIX: SHADOW CONTRAILS – THE BLACK SWAN. During recent re-analysis of weather satellite data in 2024, we discovered several shadow contrails. The phenomenon is not new, but because of its general importance for airplane detection, in particular for cases like MH370, and because of its optical similarity with the shape of a black swan neck in satellite pictures, we decided to name the phenomenon the 'Black Swan'. In the following, we describe several new contrail observations related to MH370. We focus on details of shadow contrails, but also include other observations. In principle, any airplane above the clouds during daytime hours make a shadow contrail. However, shadow contrails can also appear during the night in infrared satellite pictures.

There are two reasons for formation of shadow contrails. The first is simply the optical shadow they cast on clouds below. In most cases, this is extremely difficult to see unless conditions are good. Good conditions occur if the cloud tops below are horizontal, well defined and with smooth surface. In addition, the sun must be at relatively low inclination.

In addition, the shadow leads to local cooling of the top of the clouds. If an infrared camera is the observer, the contrast greatly improves, since there is both direct infrared optical shadow and reduced thermal emission due to lower temperature at the same time and place. The thermal part of the effect also works at night, if the contrail is under thin clouds above a warm surface (typically hot tropical sea) casting an upward heat shadow. However, the thermal effect demands one more condition. The wind speed must be low or moderate, not with too turbulent wind pattern, and predominantly along the contrail to minimize thermal mixing. We noticed this was the case several places along our proposed route for MH370 because of its location relative to the intertropical convergence zone and the related trade winds. Simultaneously, the clouds below (or in one case above) looked nice and smooth. Finally, we observed perfect shadows from some stratospheric clouds moving almost perpendicular to the normal clouds. As a first exercise, we calculated the height of these clouds. They were at 19 km above sea level in agreement with typical values. The result remained unchanged for different satellite pictures of the same cloud as long as we corrected for the changing solar inclination angle and the satellite observation angle.

This told us that Black Swans have two independent applications. They make contrail hunting easier because of better contrast. Secondly, if the corresponding normal contrail is visible, it is possible to determine the flying height above the clouds. This is valuable information, in particular for ascending or descending airplanes. However, it comes with a minor drawback. If the Black Swan is predominantly due to temperature change, it takes roughly 15 minutes to develop fully. For ideal usage, the contrail must remain stable for minimum 15 minutes, typically for 30 minutes, if standard weather satellites are used.

We started systematic contrail hunting from the northern end of our proposed flight route. Initially we thought that identification of a U-turn (final major turn for MH370?) would be most important, but we discovered much more, including proof of Kate Tee's observation. Figure 9 presents an overview of all relevant contrail features in the northern region.

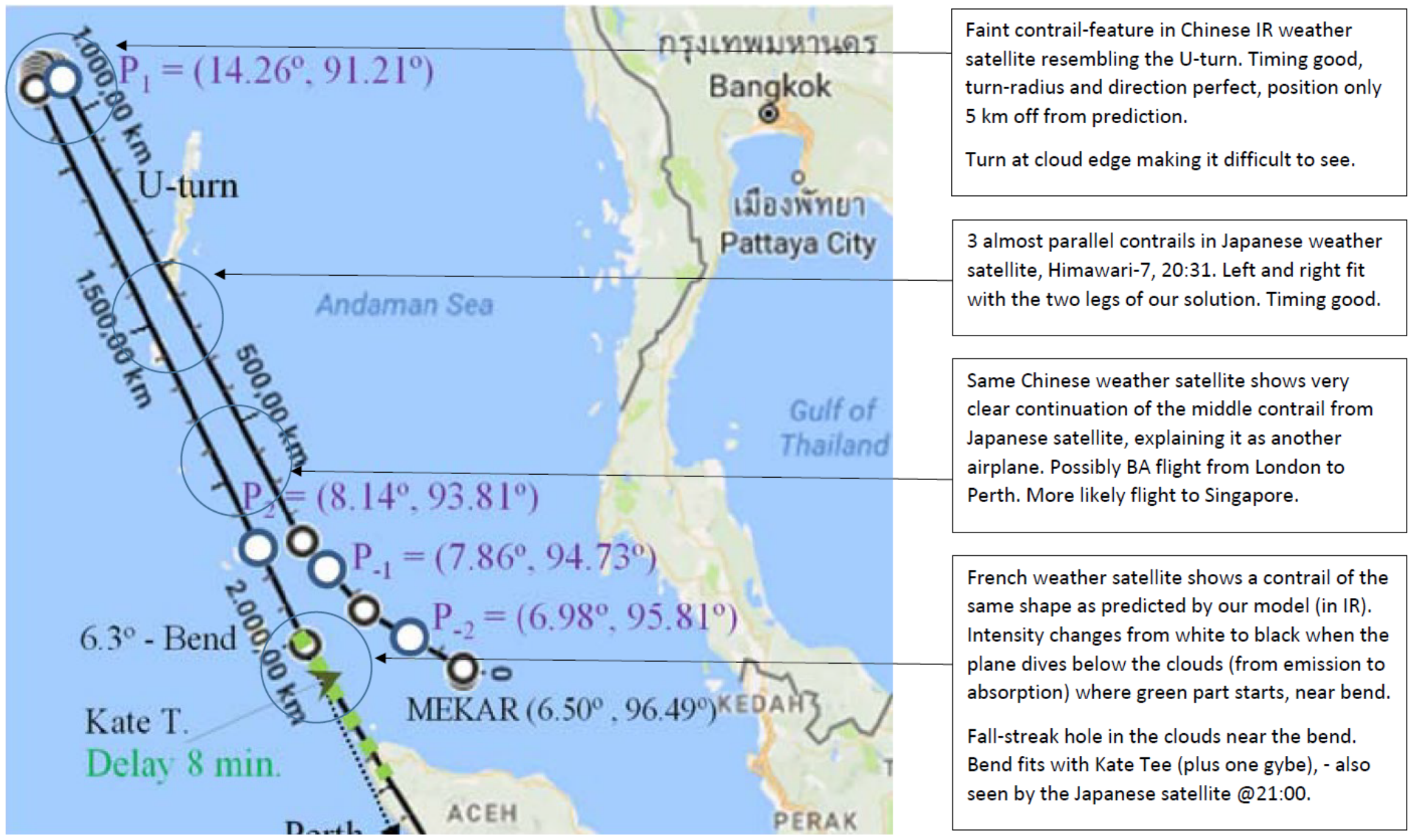

Figure 9. Overview of contrail features related to MH370 in the northern sector.

We found evidence of a U-turn at the correct place and time. A low-flying infrared Chinese weather satellite observed a U-turn along the edge of a cloud far up in the Andaman Sea. The position agrees within 5 km with our theoretical prediction of a U-turn for MH370 as shown in detail in Figure 10. It is a nice picture since the airplane is around 80 % through the U-turn, so the direction including some of the straight part leading up to the turn is visible (indicated with arrow). This kind of navigation is extremely rare under normal flying conditions over open sea, - in particular crossing an airway with risk of a collision.

Next, we found two small features in a picture from a Japanese weather satellite stationed far out over the Pacific Ocean (MTS, 2024). These confirmed the middle parts of the long trip up to and back from the U-turn. The important detail is that the angles are correct, and that a third contrail segment (between the two 'legs') clearly identifies as coming from another airplane by the same Chinese IR satellite further south.

Another satellite picture IR@22:00 from the French weather satellite Meteosat 7 over the Indian Ocean delivered extremely important information near Bandar Aceh. The previous pictures (between 19:00 and 21:30) are unfortunately absent because the satellites solar panels passed through the earth shadow, so a small correction for an hour wind drift is necessary. Slightly north of the position from where Kate Tee observed an airplane diving towards her there is a so-called fall-streak hole in the thin, hazy cloud-cover, proving that an airplane did actually dive at that position. In addition, the contrail passing through the hole 'inverts' in infrared, becoming a Black Swan. Before the hole, it is thin and white in IR (normal for a plane flying above the clouds at night), while after the hole it became thick black (clearly visible) indicating that the airplane was now below the clouds where the contrail absorbs and scatters heat from the ocean, cooling the cloud above and making the IR image turn black. When the contrail reached land in Aceh, there was no longer any cloud cover, and it turned white again. This is practically a fingerprint. The only problem is that Kate Tee got the time one gybe wrong as we already suggested in an earlier version of our paper. If her gybe-number were correct, there would not have been any trace of contrail or drop-hole left at 22:00. Instead, there might have been something in the Japanese satellite picture at 20:31, but there is nothing at that position at that time. The southbound contrail in Himawari-7 stops shortly after passing the second Andaman Island as one would expect. Until recently there was also a Japanese satellite picture available at 21:00 (unfortunately deleted before we took a copy) showing a smaller fall-streak hole and white contrail until a few km into Bandar Aceh confirming the exact time the airplane passed. This means that it could only be MH370, thereby confirming a fingerprint of the events. There is no longer any doubt Kate Tee saw MH370 flying low and slow over Bandar Aceh, but it took place one gybe later than she remembered.

As a last detail, we would like to point out that with a normal white contrail below a thin, hazy cloud at night, the visible optical contrast becomes very different, so Kate Tee thought it was smoke instead of just a normal contrail.

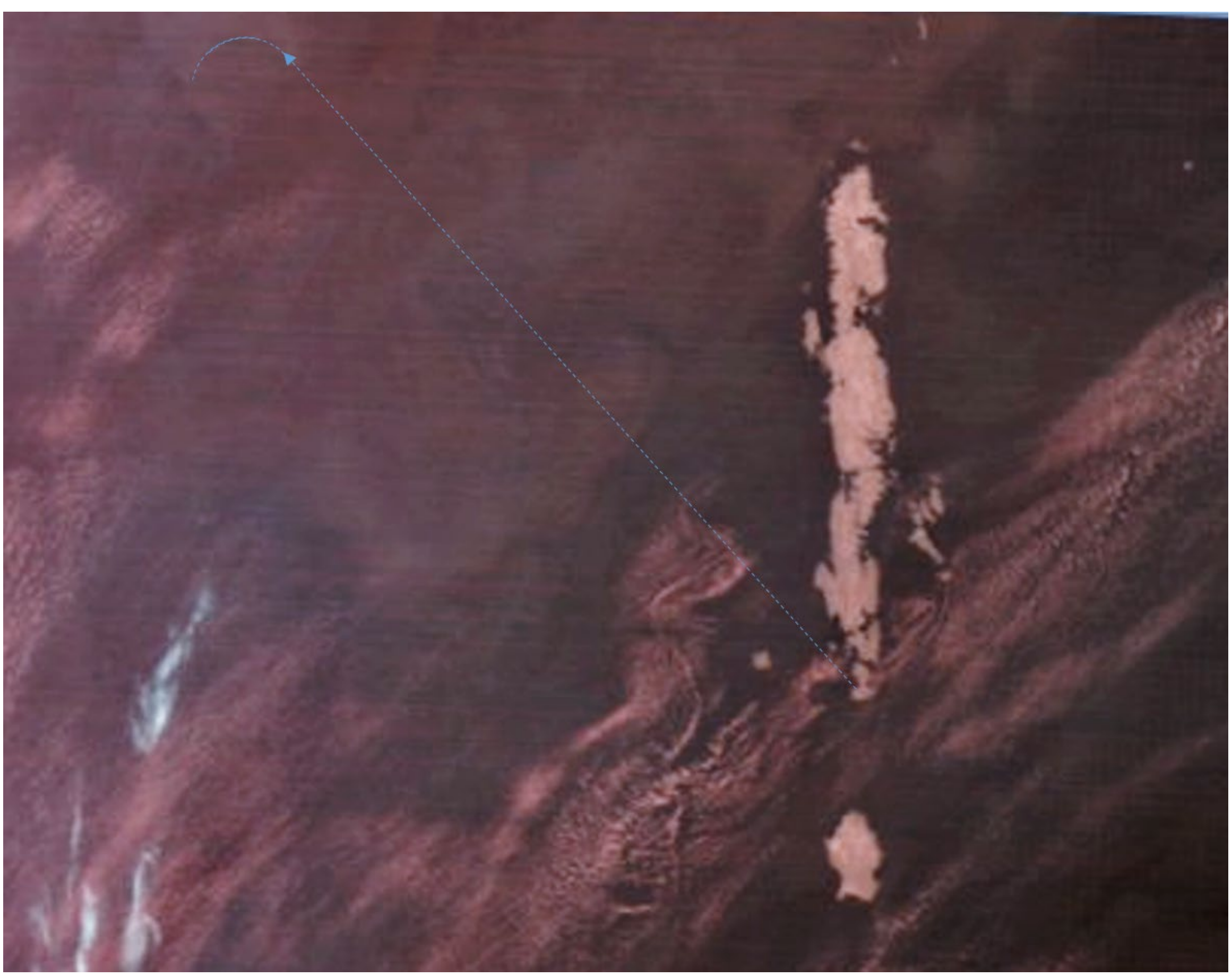

Figure 10. Chinese infrared satellite picture with the relevant part of our proposed U-turn indicated with dashed blue line in agreement with the observed contrail from shortly before until 80 % into the turn.

Figure 11 displays recorded contrails and related phenomena near Bandar Aceh observed by Meteosat 7 in infrared @ 22:00. The sequence of events displayed in the figure, including the few-degree turn (6.3 degrees in our model), agree nicely with observations made by Kate Tee, but shifted the equivalent of one gybe forward in time.

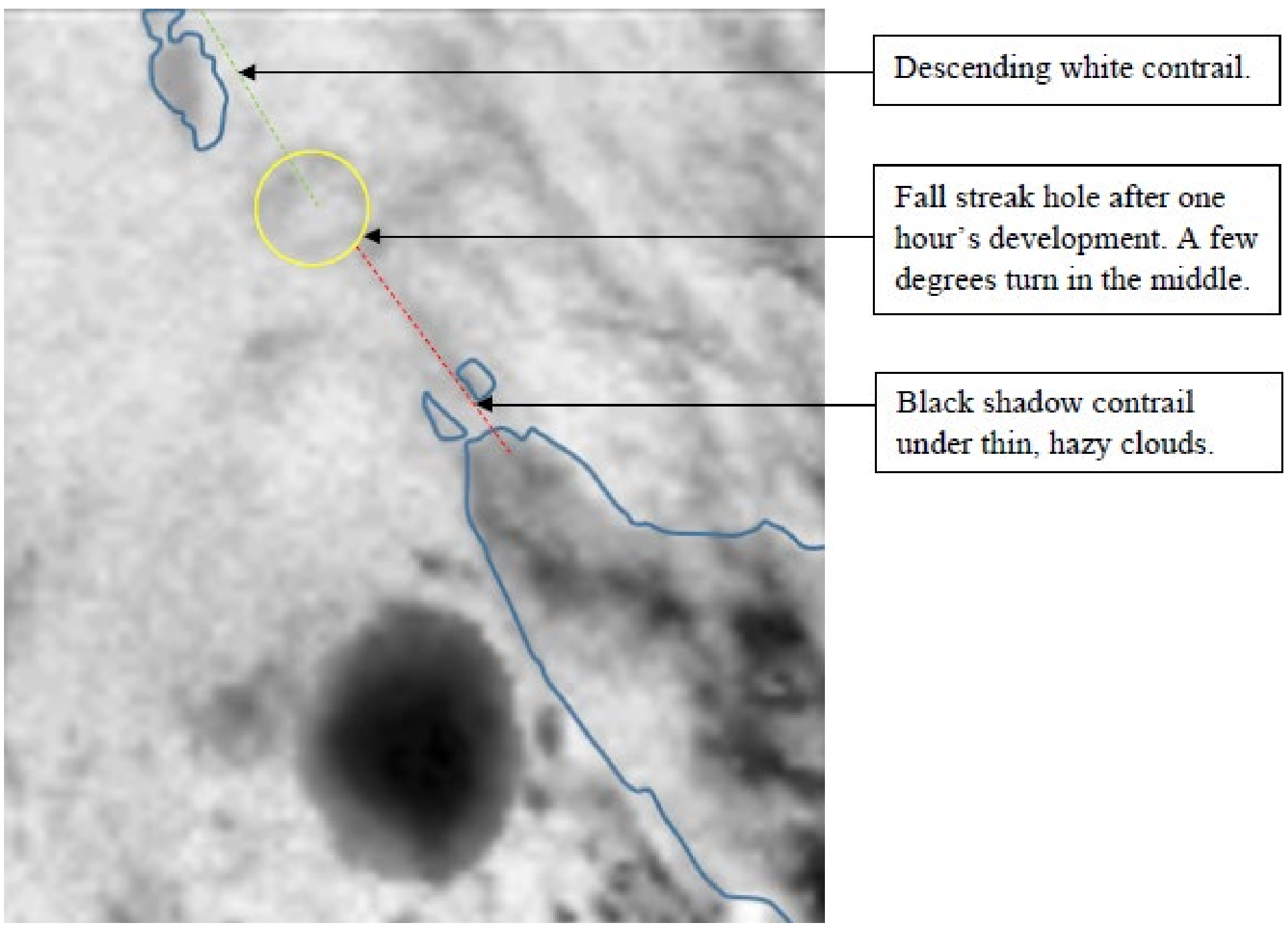

Figure 11. White contrail, fall-streak hole, and black shadow contrail observed near Bandar Aceh.

Additional information appears using an interactive display, since toggling between infrared and a near infrared (water vapour) picture show near infrared contrails shortly outside Figure 11, both above the top and below the bottom of the figure. This confirms that an airplane flew high above the clouds in both places, and dived down in between. In addition, it allows the user to determine which of the lines above the Aceh province came from the airplane contrail, and which were due to effects from wind and tall mountains. The most obviously chosen (dark-grey) line agrees with the near infrared contrails as well as with our model prediction, and the IR contrails. This confirms the whole scenario.

Finally, we are going to discuss contrail evidence from the end of the flight near the seventh arc. Several years ago, we found contrail segments near Christmas Island left by two different airplanes as shown in Figure 8. Some of it appeared so early after sunrise that shadow contrails are difficult to identify (insufficient solar heating). Fortunately, data from two seismic detectors at Christmas Island confirm the contrails and secure an extremely nice agreement with our model for MH370 for the contrail segment southwest of the island.

Figure 8 describes a presumed continuation of the contrail recorded at 00:30. No other sources confirm this part, and we are now convinced it was a fake signal generated by a data-transformation of the satellite picture to rectangular coordinates. The time and position for a heading change is correct (because the first engine ran out of fuel), but the new heading was incorrect. In 2024, we decided looking for shadow contrails to find the correct end-scenario. Figure 12 shows a clear Black Swan pointing to the opposite side with comparable angle in an untransformed IR picture from METEOSAT-7. This means the heading change did appear at the expected time and position, but the airplane turned east instead of west in agreement with the most likely electrical configuration of the airplane according to Boeing simulations. Next to the Black Swan, we also found the primary white contrail (faint).

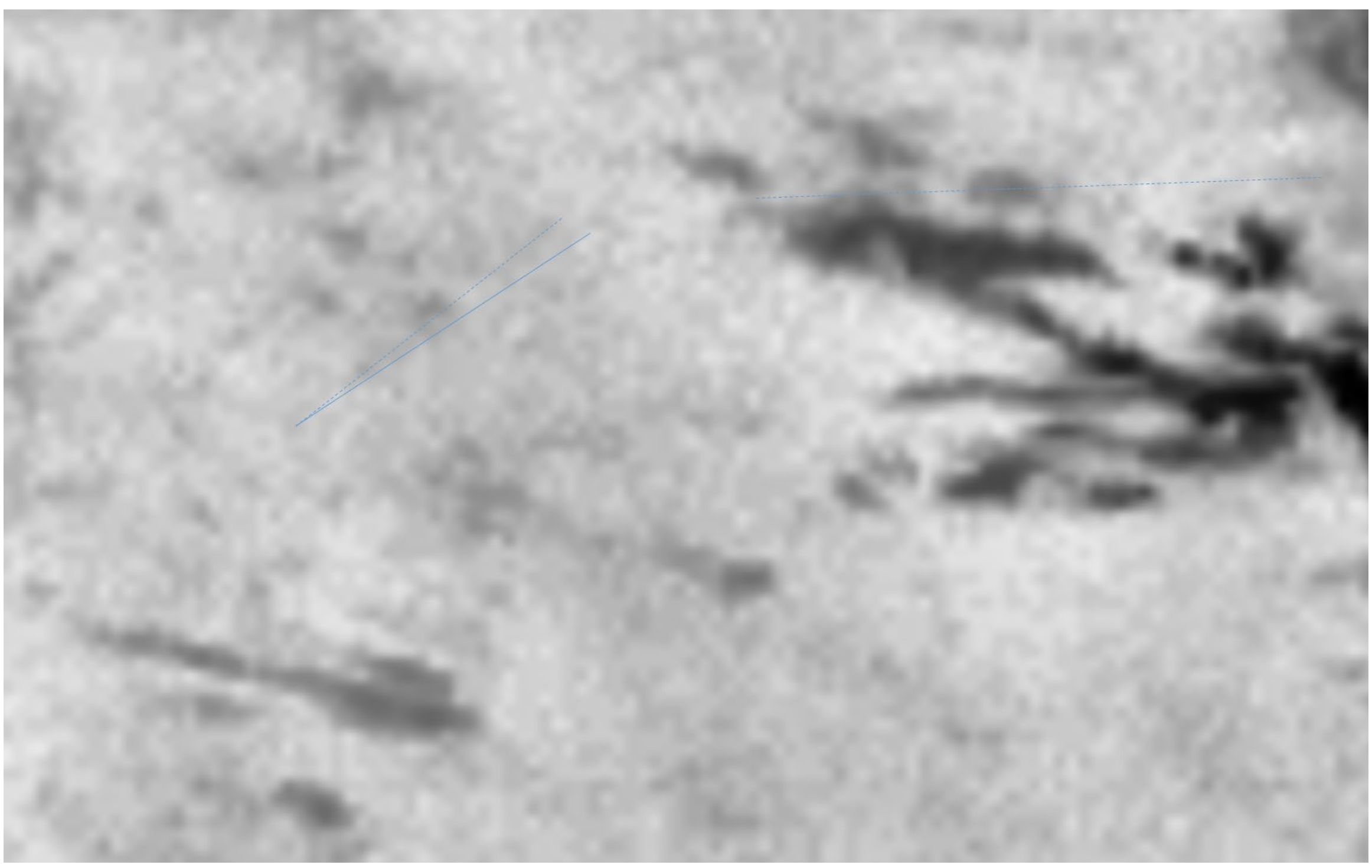

Figure 12. Shadow contrails recorded in IR at 00:30 by Meteosat-7 after Christmas Island are marked with dashed blue (notice 90° turn). A faint white contrail observed in the same picture is marked with blue line.

The normal white contrail also appears in the corresponding visible satellite picture with better contrast as shown in Figure 13. The most important detail is that the two contrails merge in one point, and no contrail is leading away from this point.

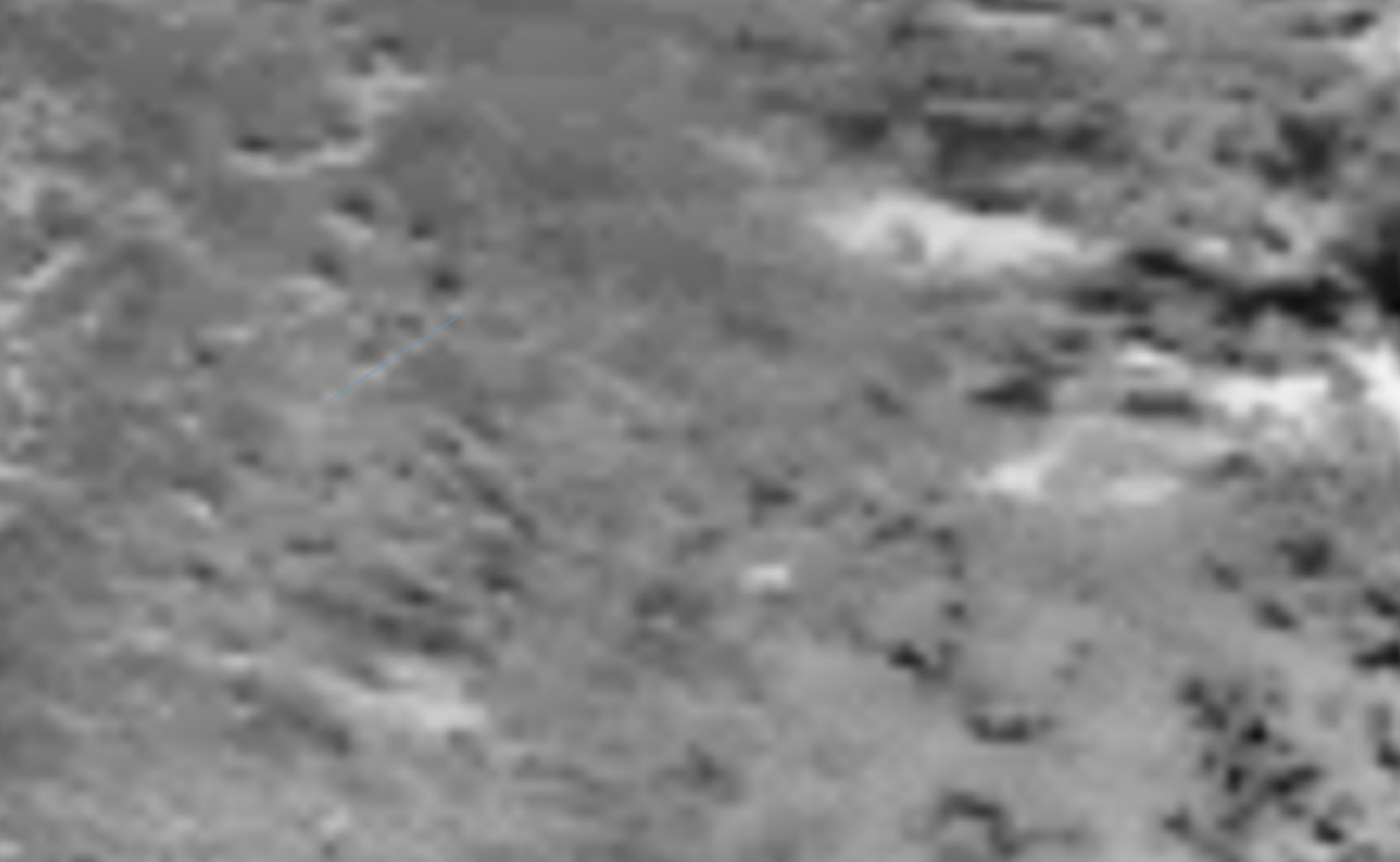

Figure 13. White contrail recorded in visible light at the same time and place marked with blue line. Notice again the 90° picture-turn and agreement with the faint white contrail in IR above, and a white feature just after the contrail. It is also visible as a black spot in IR. That is the top of a mushroom-cloud from the crash. For sharp eyes, there are spokes going from the white top to an outer ring and clouds below appear blurred. Probably the blurred ring is the shock wave from the thermobaric explosion of MH370 hitting the ocean.

Instead, we found evidence of an explosion a few km southwest of the contrail ending (in both Figures 12 and 13). The position is at (**13.53° S, 107.11° E**). In a picture from 01:00, the explosion-feature developed into a complete mushroom-cloud of the so-called Wilson type (Mushroom, 2024) with a 'crown' around its top as shown in Figure 14. This type of mushroom-cloud typically develops after underwater nuclear test explosions. We made an estimate of the total amount of energy transferred to the atmosphere during crash of an airplane like MH370. There are 3 different contributions: Kinetic energy of the airplane, energy from explosion of the lithium batteries it carried (no fuel is left, so its contribution is zero), and finally energy transferred from the hot water splashed hundreds of meters up in the couple of degrees cooler morning air. We found a total amount of energy equal to roughly 3 % of the Hiroshima bomb, explaining that a mushroom-cloud could form. The fact that it only just made it above the clouds points to a cloud top around 3 km. As further evidence, the leftover of the mushroom-cloud is still clearly visible in both infrared and visible satellite pictures recorded at 01:30, but shifted slightly north by wind drift.

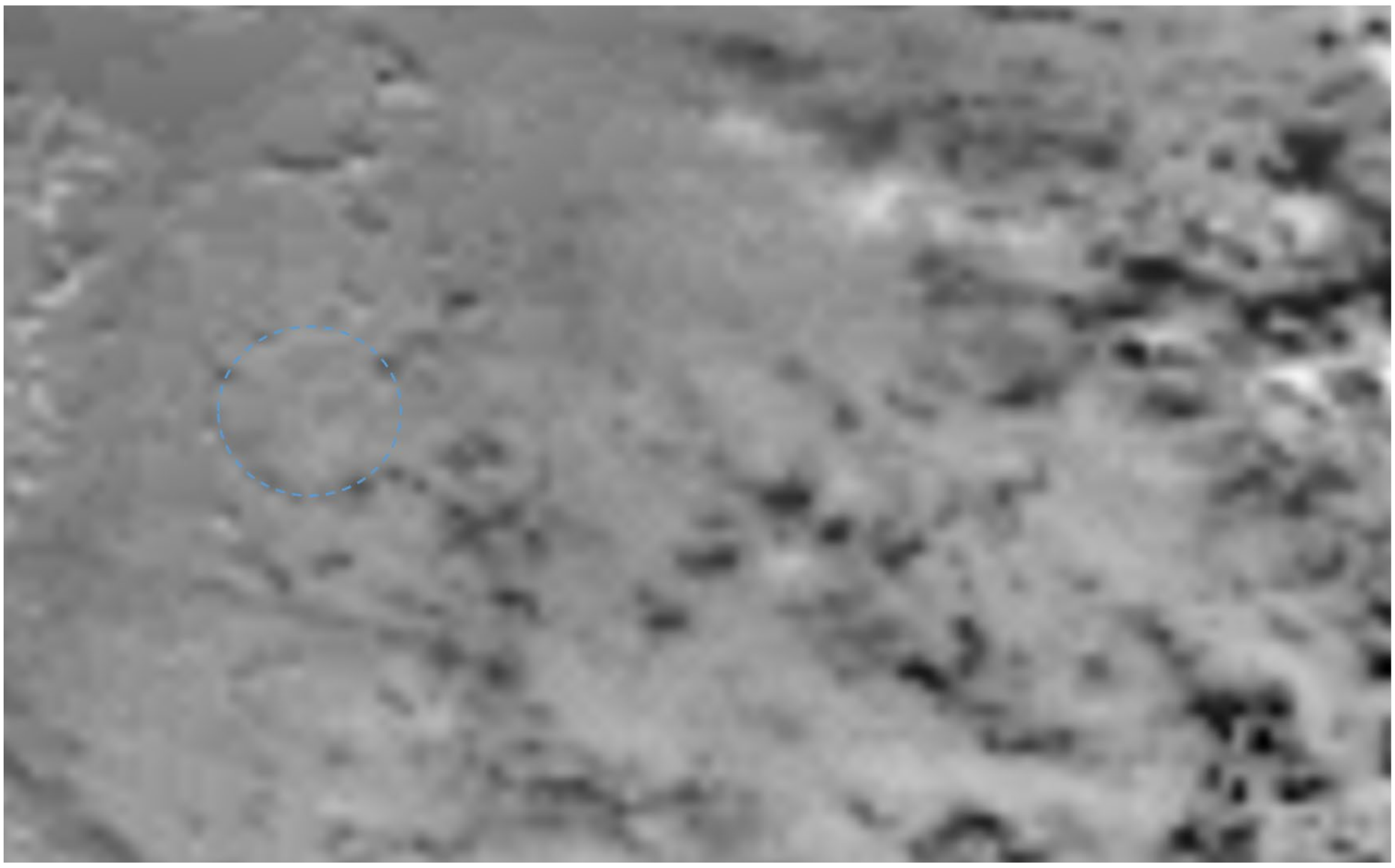

Figure 14. Visible picture at 01:00 at the same place (with a small wind correction) marked with a dashed blue ring, again 90° turned, and showing development of a Wilson cloud from the original mushroom.

With this information in mind and by including a small segment of Black Swan near Christmas Island, we can now measure the flight height during the entire end-scenario. Figure 15 displays the results together with estimates of the horizontal speed of the airplane with 10-15 % relative uncertainty. When passing closest to Christmas Island, the flying height was 10.9±1.0 km in nice agreement with our model. When the contrails merged at the end, the height had dropped to 3.0±0.5 km. This height-loss appeared within roughly 11 minutes, and simultaneously the forward velocity dropped almost a factor 4. The Holland paper tells us that soon after the end of the plot, MH370 accelerated downward by 0.68 G and developed a large vertical speed. All this evidence tell us that the airplane could only have been MH370, since no conscious pilot would have allowed such crazy navigation more than 1000 km from the nearest airport in mainland Australia. Any normal pilot would immediately after the first engine-failure have turned around and made an emergency landing in the nearest airport at Christmas Island to avoid a catastrophe. Since no other airplanes disappeared that night, it must have been MH370 loosing height and crashing.

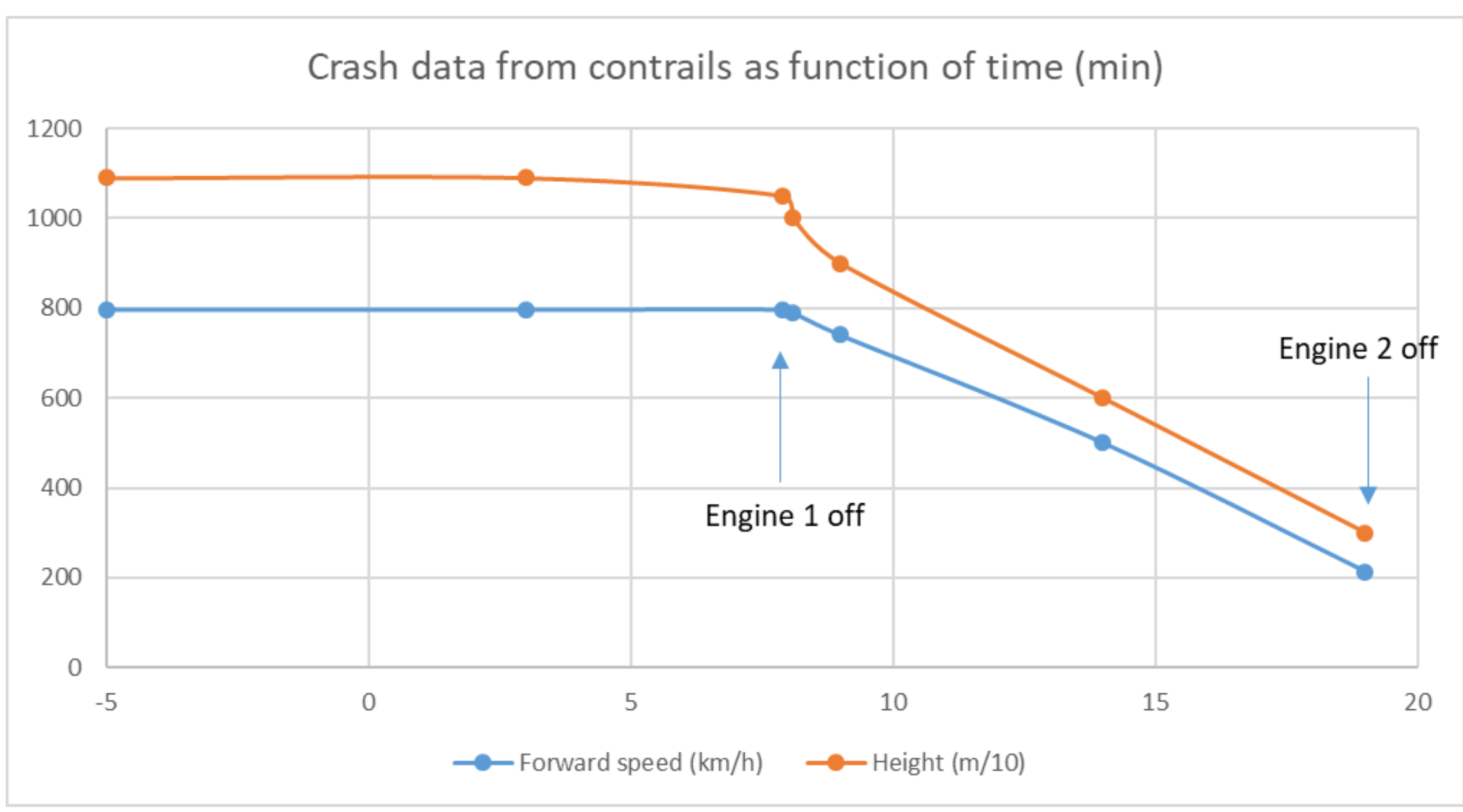

Figure 15. Measured height and estimated forward speed during the last 24 minutes of flight above the clouds.

Development of the first part of the mushroom-cloud takes only approximately 1.5 minutes, so we can practically ignore wind drift. Therefore, determination of the centre of the mushroom-cloud is so precise that the debris was within a 5 km radius of this position. However, after formation of the debris-field, wind and currents in the ocean can move things a few km horizontally before they reach the seabed. We therefore recommend searching a 10-km radius around (**13.53° South, 107.11° East**) to find MH370.